# Heat and Worker Health

Andrew Ireland, David Johnston, Rachel Knott

Centre for Health Economics, Monash Business School

June 1, 2023

Extreme heat negatively impacts cognition, learning, and task performance. With increasing global temperatures, workers may therefore be at increased risk of work-related injuries and illness. This study estimates the effects of temperature on worker health using records spanning 1985–2020 from an Australian mandatory insurance scheme. High temperatures are found to cause significantly more claims, particularly among manual workers in outdoor-based industries. These adverse effects have not diminished across time, with the largest effect observed for the 2015–2020 period, indicating increasing vulnerability to heat. Within occupations, the workers most adversely affected by heat are female, older-aged and higher-earning. Finally, results from firm-level panel analyses show that the percentage increase in claims on hot days is largest at 'safer' firms.

___

Corresponding author: Andrew.Ireland@monash.edu; Level 5, Building H, Monash University Caulfield, Victoria 3145, Australia
We thank seminar participants at Monash University and conference participants at the 11th Australasian Workshop on Econometrics and Health Economics and the UCLA Climate Adaptation Research Symposium for discussions and feedback. This research uses unit record data from the Compensation Research Database (CRD). The CRD contains administrative data from WorkSafe Victoria however the findings and views reported in this publication are those of the authors and should not be attributed to WorkSafe Victoria.

# 1 Introduction

The rise of global average temperatures over the past century has increased workers' exposure to extreme heat. Given that high temperatures can impact task performance (Park, 2022; Seppanen et al., 2006), learning (Park et al., 2020) and cognition (Carias et al., 2022; Heyes & Saberian, 2019; Künn et al., 2019), increased heat exposure implies that workers may have a higher risk of experiencing workplace injuries and illnesses. Workplace-related injuries and illnesses can lead to significant medical expenses, reductions in long-term earnings and productivity, and higher business costs (e.g. recruitment and training of substitute workers), implying that the total cost of worsening occupational health could be substantial. Increases in workplace injuries and illnesses may also exacerbate economic inequalities if increases in risk are concentrated among lower-earning workers (Park et al., 2021). Therefore, it is important to understand the types of workers who are most affected by high temperatures, how worker health is affected (e.g. types and severity of injuries and illnesses), the importance of firm-level factors, and whether there is evidence of adaptation to higher temperatures (e.g. through workplace heat policies and changes in technology). Such understanding will additionally be important for identifying and quantifying the full extent of damages due to anthropogenic climate change, and the development of climate change mitigation programs and policies. Importantly, the magnitude of any heat-accident relationship may be context-specific, depending upon the location, time-period, and industrial relations and mitigation contexts. Consequently, analysis of data from different settings is required to gain a full understanding of the impact of heat on workplace health.

In this study, we explore the causal effects of outside ambient air temperature on the incidence and nature of occupational health claims using a large administrative database. Our data contains over two million occupational health claims submitted to a state-run mandatory workers' compensation scheme in the Australian state of Victoria from 1985 to 2020.[1] Victoria is an attractive jurisdiction for this analysis because its climate is highly varied, with daily maximum temperatures often varying over 20°C within a location and month. The large variation enables an estimation approach that compares differences in accident rates across days within the same postcode and month. Second, workers' compensation insurance has been mandatory in Victoria since 1985, meaning we can analyse differences in estimated effects across a 35-year period. Third, some Victorian workplaces have implemented heat policies due to extreme summer temperatures; for example, construction workers often stop working in extreme heat. The effects we measure are the residual risk after accounting for such adaptation measures and therefore provide evidence on the likely effects if other jurisdictions implement similar policies.

---

[1] Victoria is the second most populous state in Australia, with a labour force of 3,448,100 employed persons as of February 2020 (Australian Bureau of Statistics, 2020a).

We find that higher maximum temperatures generate significantly more claims. A one-degree Celsius increase in the daily maximum increases claims by 0.24%, or alternatively, a day reaching 33–36°C averages 5.3% more claims than a day 18–21°C. The adverse impacts of high temperatures are observed throughout the 35 years despite considerable economic and heat-related policy changes, with the largest effect observed for the 2015–2020 period. We find that labourers (such as cleaners, farm hands and kitchenhands) and tradespersons (such as metal fitters, carpenters and motor mechanics) are particularly vulnerable to the impacts of heat and that the effects are largest for workers in outdoor industries like agriculture and construction. We also find differences in effects by worker characteristics. Within manual occupations, estimates are largest for female, older-aged, and high-earning workers.

Subsequent analyses reveal that the impact of high temperature is not limited to heat stress and sunburn, with various health events, such as lacerations and other injuries, occurring more frequently on hot days. This suggests that reduced concentration and psychomotor performance are drivers of the reported effects. In addition, we find that the risk of accidents caused by environmental agents, such as hazardous chemicals and substances, is also heightened on hot days.

Regarding workplace characteristics, we find that workplaces with small- and medium-sized staff numbers are more impacted by heat. A possible explanation is that large workplaces, which tend to have higher unionisation rates and where collective bargaining with employers is possible, are more likely to adopt heat thresholds for the cessation of work, which reduces the harmful effects of extreme temperatures (CFMEU Victoria, 2015). We also analyse a panel of firms, matched on their industry and size, and find that the percentage increase in claims on hot days is largest at "low claim" firms (which we use as a proxy for general safety levels).

This study contributes to recent studies on the causal impacts of heat on worker's compensation claims (Dillender, 2021; Filomena & Picchio, 2022; Page & Sheppard, 2019; Park et al., 2021).[2] Our main estimates demonstrate that the adverse effects of temperature on work-related accidents can be replicated outside the U.S., despite differences in labour market conditions, occupational health and safety systems, union coverage, and climate. Most similar to our study is Park et al. (2021), which uses an occupational claims database from California, where workers' compensation insurance is also mandatory. In contrast with Park et al. (2021), we do not find larger heat effects amongst low-income workers using claimant-level pre-injury earnings. Within high-risk occupations and industries, we find that medium- and high-earning workers are most sensitive to heat.

---

[2] There are also several studies in the public health literature concerned with the safety of workers on hot weather that establish a positive correlation between high temperature and occupational health claims. These studies are typically limited to examining variation in temperature in a single location over summer months. Studies in an Australian context include Xiang et al. (2014) and Varghese (2019) which both associate high temperatures with increased occupational health claims. While informative, these studies are unable to tell us about the extent to which heat causes adverse worker outcomes. In particular, the observed correlations could be confounded by other factors that affect worker's health and safety which are also correlated with temperature (e.g. selection into certain jobs).

Our results also contribute to the discussion regarding workers' ability to adapt over time. Dillender (2021) proposes there are limits on workers' ability to adapt, while Park et al. (2021) find a significant reduction in heat-related injuries between 2000 and 2008, suggesting that adaptation likely occurred over this period. We find that heat effects amongst high-risk occupations have remained surprisingly stable over the past decades, with the largest effect occurring in the 2015–2020 period. Our results indicate that adaptation is difficult for workers in outdoor settings, consistent with Dillender (2021). More broadly, our research contributes to a wider literature concerned with the impact of global climate change on workers' mortality (Sarmiento & Longden, 2020), productivity (Zhang et al., 2018) and time use (Graff Zivin & Neidell, 2014).

The paper is organised in the following structure. Section 2 gives a brief overview of Victoria's workers' compensation scheme. Section 3 and Section 4 describe the data sources and empirical approach. Section 5 provides estimates of the overall effect of temperature on the number of claims and the effect by occupation over time. Section 6 focuses on heterogeneity in the effects of temperature by worker characteristics and accident types. Section 7 investigates whether firm's risk management strategies can mitigate the effects of heat. Section 8 includes back-of-the-envelope calculations to estimate the cost of heat-related accidents. Finally, the study concludes with a discussion on workers most vulnerable to working in heat and effective strategies for adaptation.

## 2 Workers' Compensation Scheme in Victoria

Victoria introduced a mandatory workers' compensation scheme in September 1985 with the establishment of WorkCare. Firms pay premiums to cover the costs of workers' injuries and illnesses arising because of work. The scheme covers approximately 85 percent of the labour force in Victoria; sole traders can be exempt from registering, federal government employees are covered under commonwealth compensation schemes, and approved self-insured companies manage their own claims processes (Prang et al., 2016; WorkSafe Victoria, 2021b).[3]

There are several types of compensation available to workers in Victoria. Claims can include compensation of medical expenses and a weekly payment based on their pre-injury average earnings for their time away from work. Workers and their dependant partners can also be entitled to lump sum payments in permanent disability or death cases. The scheme operates under the "no fault" principle, meaning that workers are not obliged to demonstrate the firm was responsible to receive compensation. Workers and employers must submit a medical certificate as evidence to lodge a claim, which is assessed by agents who determine payments for medical treatments and time off work. The employer is liable to cover the first 10 days of weekly benefits for time off work and the first AU$735 of medical expenses as an excess on accepted claims.[4]

The scheme has transformed since it was first introduced (see Appendix A for a summary). The most significant changes occurred in 1992 when the election of a new government resulted in a system re-design to reduce costs (see further discussion and time series data in the Data section). In this study, identifying variation comes from differences in the maximum daily temperature within a postcode within a month-year. This variation is unlikely to be affected by changes to the scheme. However, we present estimates separately for different periods, allowing for heterogeneity in the effects of temperature across time. Note also that we present results as a semi-elasticity (i.e. the percentage change in the number of claims for each 1°C) to aid comparisons between different periods.

## 3 Data

### 3.1 Occupational Health Data

We source occupational health claims from the Compensation Research Database (CRD), which records all claims submitted to Victoria's mandatory workers' compensation insurance scheme since its inception in 1985. The CRD recorded 2,156,960 claims corresponding to 1,159,249 unique claimants between 2nd September 1985 and 31st July 2020, with all dates referring to the date of affliction. A summary of claims is provided in Table 1. Males account for 70% of claims overall and 83% of claims for labourers and tradespersons (high-risk occupations we discuss throughout the paper). We divide the sample into pre- and post-2000 in subsequent analysis. Labourers and tradespersons represent 418,165 out of the 1,026,096 claims since 2000 (41%).[5] In our heterogeneity analysis, we estimate the effects separately by industry. Outdoor-based industries, which we expect to be most temperature sensitive, such as 'Agriculture, Forestry and Fishing' and 'Construction' account for 3% and 19% of claims for labourers and tradespersons since 2000 respectively.

---

[3] As of December 2021, there were a total of 34 self-insuring firms (which is higher than in earlier years). WorkSafe Victoria assesses and grants approval to prospective self-insurers based on their financial strength, resources to administer claims, and their safety record. The self-insured firms include retail, energy, food and financial corporations. It is possible that self-insured firms are safer, however their exclusion is unlikely to have a meaning impact on our results since they account for a small percentage of the labour force.
[4] Medical excess was AU$250 in 1985 indexed annually increasing to SU$735 in 2021. Employers were only liable for the first 5 days between September 1985 and December 1992. There is an option for employers to remove their liability of paying the excess by paying 10% additional in their premium (WorkSafe Victoria, 2021a).

The most disaggregated level of geography provided in the CRD is postcodes. Therefore, we use the workplace's postcode as the geographic variable to match climatic variables. All postcodes with claims were included in the analysis, equating to 694 out of 698 postcodes. The median postcode is 93km$^2$ and contains approximately 5,000 employed persons (Australian Bureau of Statistics, 2020a).

Demographic information about the claimant, such as gender, age, occupation, and industry, are included for each claim. A unique feature of the database is that each workplace is given an identifier making it possible to study heterogeneity between high-claim and low-claim firms, which we do in Section 7. The database also includes sufficient detail to classify the nature of the affliction (e.g. burn, fracture), accident type (e.g. fall, hitting object), accident cause (e.g. powered equipment, chemicals) and primary bodily location of affliction (e.g. neck, lower limb). The pre-accident earnings of the claimant are also recorded, enabling exploration of heterogeneity in the effect of heat by earnings. Previous studies have relied upon occupation- and area-level aggregate measures of income (Park et al., 2021).

The number of monthly claims has decreased over the analysis period (Figure 1), with a sharp decrease in December 1992.[6] This corresponds to legislative changes that restricted the benefits paid to workers to improve the scheme's financial viability (Accident Compensation (WorkCover) Act 1992 (Vic), 1992; Stylianou, 2011). Claims subsequently increased in the following years as the scheme achieved financial stability, which first became fully funded in the 1994/95 financial year. Volatility in the number of claims due to changes in the scheme may complicate comparisons in the heat effect over time, which is discussed further in Section 5.2. Since 2000, the period we focus on for much of the analysis, there have been relatively few changes to the scheme and the number of claims has steadily declined as a result of improvements in workplace safety controls, increased awareness of workplace health and safety, and reduced exposure to hazards from a shift in the composition of the labour force away from high-risk manual occupations (Australian Bureau of Statistics, 2020b; Safe Work Australia, 2020). There is also seasonality in the number of claims, with fewer claims in December and January. This is expected because many workers are on holiday during these Australian summer months.

---

[5] Multiple claims for a single claimant (e.g. when a second claim made for additional medical expenses not manifested at the time of the original claim) with the same date of affliction are counted as one when tallying the daily number of accidents. The study includes all claims, regardless of whether the claim was accepted by the insurer. In a robustness analyses we also test the sensitivity of our results to using only accepted claims (Appendix E).
[6] Claims may take several years to be reported after the affliction, however, the majority are reported within the same year. As at May 2021, 80% of all claims reported relating to afflictions in 2010 were reported in 2010 (95% reported in 2010 and 2011).

On average, there were 209 claims per day in Victoria. We use the daily number of claims in a postcode as our main outcome variable of interest, which has a mean of 0.54. The distribution for this variable is positively skewed, with 71% of postcode-days having zero claims and 1.7% of postcode-days having 5 or more claims (see Appendix B).

*3.2 Climatic Data*

Daily weather observations from 73 weather stations and 1,101 rainfall-only stations were obtained from the Bureau of Meteorology. To interpolate observations from weather stations to postcodes we use a technique called residual inverse distance weighting. Spatial interpolations methods based on regression residuals (Ninyerola et al., 2007; Wu & Li, 2013), particularly when including elevation (Stahl et al., 2006) and distance to coast (Salet, 2009), have been shown to be a superior measure of temperature (Nalder & Wein, 1998). First, using observations at weather stations where the temperature is known, we use Ordinary Least Squares (OLS) regression with maximum temperature as the outcome variable to obtain coefficient estimates for latitude, longitude, elevation and distance to coast as regressors. We also calculate residuals from this regression using the observed maximum temperature value. Our predictions for the maximum temperature at the postcode centroid are calculated as the sum of: (i) a "deterministic" component using the OLS coefficients and the latitude, longitude, elevation and distance to coast of the postcode centroid; and (ii) a "stochastic" component by interpolating the residuals using inverse distance weighting. Previous studies use measurements from the closest station (Park et al., 2020), inverse distance weighted average from nearby stations (Zhang et al., 2018), and limit the sample to smaller regions (Xiang et al., 2014). Using cross-validation, we demonstrate that residual inverse distance weighting has the lowest root mean squared error compared to these alternatives, making it the preferred choice to reduce measurement error. The results of cross-validation are provided in Appendix C.

Victoria has a varied climate, ranging from a hot semi-arid climate in the North West to cold mountain climates in the Great Dividing Range. Most of the state, including Melbourne (Victoria's capital city), experiences a temperate climate with a warm summer from December to February.[7] The number of days over 30°C and 35°C each year has slightly increased during the analysis period, with 34 days > 30°C and 14 days > 35°C in Melbourne during 2019, our last complete analysis sample year (see Appendix D). Furthermore, within-month climatic variation in a postcode, which is used for identification, is relatively large. In Melbourne, the average difference between the highest and lowest maximum temperatures in a month is 19°C in summer and 10°C in winter. We do not assess the effect of extreme cold temperatures, which may also plausibly harm workers, since temperatures do not regularly get as low in Victoria as they do in many other parts of the world.

## 4 Empirical Approach

We employ an area fixed-effects regression model to estimate the causal effect of outside ambient air temperature on the daily number of occupational health claims. We model the number of claims in each postcode ($p$) for each day ($t$) as a Poisson-distributed random variable as shown in Equation (1).[8]

$$E[claims_{pt}|X] = \exp(\beta * maxtemp_{pt} + \phi_{pm} + \gamma * X_{pt}) \quad (1)$$

The preferred specification includes month-year-postcode fixed-effects, $\phi_{pm}$, to allow for time-specific factors to differ across the 694 postcodes within the state of Victoria. In this specification, the impact of heat is identified through day-to-day variation in temperature within a given month-year of a specific postal-area. Under the assumption that temperature varies randomly over time within a given local area, which is recognised as a plausible assumption in the climate-economics literature (Dell et al., 2014; S. Hsiang, 2016), estimates of $\beta$ can be interpreted as causal. Month-year-postcode fixed-effects also account for the seasonal association between heat and labour supply.

Regression covariates, $X_{pt}$, include day of week dummy variables and precipitation (mm). Day of week variables are included to control for differences in hours worked and composition of industries working on each day of the week. Precipitation is included because it is positively associated with injury risk and negatively associated with temperature.

---

[7] Much of Victoria's climate is comparable to other warm temperate regions such as Rome (Italy), Oakland (USA), Kunming (China) and Madrid (Spain). The average number of days over 30°C in Melbourne is expected to increase from 31 days per year during the analysis period to 43 by the end of the century (CSIRO and Bureau of Meteorology, 2022). Regions with this number of high temperature days per year include Adelaide (Australia) and Jerusalem (Israel).
[8] Estimation completed using maximum likelihood (Correia et al., 2020). Fixed-effects Poisson regression produces consistent estimates even if the claims data does not follow a Poisson distribution (Wooldridge, 1999), including data with a large number of zeros in the dependant variable (Blackburn, 2015).

Our primary estimation sample omits postcode-months with zero claims every day of the month. We apply this restriction to ensure the sample is the same across all regression specifications.[9] We also restrict the sample to weekdays, which includes approximately 90% of claims. Although we control for day of the week, we expect the temperature to have a different effect on claims during weekends due to the smaller and different population of workers. We show in Appendix E that the results are robust to this restriction. In total the sample contains 3,533,474 observations of the daily count of claims in a postcode.

We also present estimates using alternative regression specifications. Our estimates are robust to including calendar date (i.e. Jan 1st, Jan 2nd) fixed-effects, similar to the primary specification used by Dillender (2021).[10] This specification controls for seasonality, accounting for factors such as public holidays or end-of-month effects that may influence the daily number of claims in a postcode. We further demonstrate robustness by using a specification with separate month-year and month-postcode fixed-effects, which aligns with the main specification used by Park et al. (2021).

In robustness analyses, we consider the inclusion of controls for air pollution (PM2.5), which has been shown to negatively impact workers' health (Raaschou-Nielsen et al., 1995) and productivity (He et al., 2019); and is positively associated with temperature (Kalisa et al., 2018; Li et al., 2019). We do not include air pollution in our preferred specification because the amount of industrial activity, which may be affected by accidents, contributes to the amount of air pollution. Also, the most accurate measurements available for air quality for the entire analysis period are from the MERRA-2 reanalysis data, which likely relied on temperature inputs.[11]

We do not observe the number of people working each day (i.e. the number of at-risk workers). This is difficult to measure for a whole workforce and also varies with temperature (Graff Zivin & Neidell, 2014; Somanathan et al., 2021). Instead, we present results as a semi-elasticity, as described below, and should be interpreted as the residual effect of temperature on worker health net of adaption measures, such as reducing labour supply.

---

[9] Postcode-days with no claims are included in the estimation sample (71% of postcode-days in our sample have no claims). However, postcode-months with zero claims on every day of the month are omitted because there is no within postcode-month variation to model.
[10] Dillender (2021) includes day-level fixed effects such that the effect of temperature is identified from temperature differences across regions on the same day.
[11] Insufficient documentation is available to confirm the measurement procedure for air pollution to rule out the possibility that temperature is an input variable.

There are several limitations to using occupational health claims as a measure of worker health. First, not all work-related illnesses and accidents result in claims, particularly minor events when the injury cost is below payment thresholds. We do not observe if heat affects minor accidents in a similar way to major accidents that result in a claim. Thus, our estimates should be interpreted as the effect on more serious accidents. Importantly, our data has thorough coverage of serious accidents since workers' compensation is mandatory in Victoria. Second, illnesses caused by long-term exposure to heat aren't detected as they are not easily attributed to a single date of affliction. There is also the possibility of fraudulent claims, however, these are believed to be rare and unlikely to impact our results, particularly after the introduction of standardised medical certificates in 1987.

## 5 Temperature and Workers' Health

*5.1 Main Effects of Temperature on Workers' Health*

Table 2 shows the estimated effects of maximum daily temperature on occupational health claims using various regression specifications, with standard errors clustered at the postcode level in parentheses. For each regression, the estimated temperature effect is presented as a semi-elasticity, which is interpreted as the percentage change in the number of claims for a 1°C increase in maximum temperature. Column (1) presents our main specification. The semi-elasticity of 0.24 indicates that a 1°C increase in maximum temperature results in 0.24% more claims or an additional 0.5 claims in Victoria per day.[12] Estimates are similar across alternative specifications, which include additional controls for air pollution (column 2), calendar dates (column 3), and month-year fixed effects and postcode-month fixed effects, rather than month-year-postcode fixed effects (column 4). The similarity of the estimates demonstrates that our main estimated causal effect of temperature on occupational health claims is robust to using alternative specifications seen within the economics literature.

Results from additional robustness tests are shown in Appendix E. First, we show that estimates are insensitive to using only claims accepted by the insurer (Table E1 and Figure E1). This supports our assumption that the extra claims on hot days result from real impacts to health. Second, we demonstrate robustness to using Ordinary Least Squares (OLS) regression instead of Poisson regression (Table E1 and Figure E2). We also verified when estimating using OLS that adjusting standard errors to account for contemporaneous spatial correlation between postcodes does not markedly alter them (Conley, 1999; S. M. Hsiang, 2010). Third, we limit the analysis to days without rainfall (Table E1 and Figure E3) to account for the possibility that workers are unable to perform dangerous activities on rainy days, which are more likely to occur during cooler temperature. Estimates for dry days only are similar to the main analysis sample. Finally, we estimate our model including weekend days, which has little impact on results because most (90%) accidents occur on weekdays (Table E1 and Figure E4).

---

[12] The marginal effect is calculated using the daily average number of claims in Victoria and multiplying by the semi-elasticity (i.e. 209 * 0.24/100=0.50).

*5.2 Has there been Adaptation to Heat Over Time?*

We expect that heat may have a larger effect on workers based on the type of work and the work environment. In particular, manual workers in outdoor-based settings have been shown to be more vulnerable to extreme heat (Dillender, 2021; Page & Sheppard, 2019; Park et al., 2021). We estimate the effect of temperature separately by occupation, both prior- and post-2000, which approximately divides the sample in half (Table 3).

The results in Table 3 show that heat has increased work-related accidents both prior- and post-2000, with each additional 1°C resulting in 0.27% and 0.20% additional claims, respectively. However, this reduction is more subdued when looking within occupations. In particular, the effect of temperature for labourers and tradespersons has only slightly decreased (from 0.36% to 0.34%). This suggests that a compositional change away from high-risk occupations likely drives the overall reduced effect. Indeed, the proportion of Victoria's labour force who work as labourers, tradespersons and related workers has reduced from approximately 32% to 23% of employed persons during the analysis period (Australian Bureau of Statistics, 2020b).

We identify tradespersons and labourers as high-risk occupations, consistent with our prior expectations. Tradespersons includes occupations such as motor mechanics, plumbers and carpenters, and labourers includes other manual workers such as cleaners, farm hands and process workers. For tradespersons and labourers, an increase of 1°C in the maximum temperature leads to a 0.35% increase in claims. The estimated effect for labourers and tradespersons is significantly larger than the 0.12% estimate for all other occupations ($p<0.01$) and the 0.16% estimate for production and transport workers ($p<0.01$), the occupation group with the next largest effect. These high-risk occupation groups have a high number of claims overall, accounting for 49% of claims but only 26% of the labour force (Australian Bureau of Statistics, 2020b), suggesting that heat exacerbates the risks in already high-risk workplaces.[13]

To further investigate trends over time, Figure 2 displays estimates for labourers and tradespersons in five-year time periods. We estimate the effects for only labourers and tradespersons to account for changes in labour force composition. The effects are positive and significant ($p<0.01$) for all year groups, indicating that hot weather has remained a significant threat to manual workers during the past three decades, despite policy attempts to address the risk and changes to the scheme in the 1980s and 1990s. Notably, the most recent five years has the largest estimate (0.47% per 1°C), although the difference in the effect from other years jointly is not statistically significant ($p=0.19$). These five years were also the hottest five years on record, both globally (GISTEMP Team, 2021) and in Australia (Bureau of Meteorology, 2021). The largest effect in the warmest time period may imply that there are limits on workers' ability to adapt. Adaptation measures, such as altering work hours to avoid hot days, may become increasingly difficult in warmer climates. This is consistent with Dillender (2021) who finds larger temperature effects in warmer climates using data from the mining industry in the United States.

*5.3 Shape of the heat-accidents relationship*

Up to this point, we have included temperature as a linear term in our regression. To allow for nonlinearity in the effect of temperature, we estimate a regression with ten 3°C bins for the daily maximum temperature. Figure 3 shows the estimates for each bin using 18–21°C as the benchmark for all occupations (Panel A) and labourers and tradespersons (Panel B) in the post-2000 time period. In both panels, the heat-claims relationship appears to be approximately linear for much of the temperature distribution. Labourers and tradespersons make 6.2% additional claims when the maximum temperature is 33–36°C, compared to 18–21°C. A notable exception is the estimate for temperatures above 36°C, which is significantly lower than for days 33–36°C for both panels (A: $p<0.01$; B: $p=0.02$). There are on average 4.9 days per postcode-year above 36°C.

We find that for much of the maximum temperature distribution, our estimated effect sizes align with studies using data from Texas (Dillender, 2021) and California (Park et al., 2021). Days reaching 31°C have between 4.4% and 5% additional claims than the reference category in all three jurisdictions. This is despite the considerable difference in economies, climates, occupational health policies, and workers' compensation systems. The shape of the relationship is almost monotonically increasing in all three studies, and we don't observe an increase in accidents in cold temperatures. Hence, a warming in climate would only adversely impact workers' health for these climates. However, there is the possibility that the risk may increase in extreme cold temperatures below those experienced in Victoria, Texas and California.

---

[13] We also acknowledge that workers may select into occupations based on their perceived and actual ability to tolerate working in high temperatures. We believe that the likely effect of such selection would be that workers with higher tolerance to heat would disproportionately work in higher risk settings. As such, the larger effect of heat for labourers and tradespersons would likely represent a conservative estimate for the true difference in risk between occupations.

A notable difference between studies is the smaller estimated effect in Victoria at extreme temperatures (>36°C), which we demonstrate in Appendix Figure G1. The results from Texas reveal large effect for the hottest temperature range, and the results from California show a modest decline in effect, but only for temperatures over 42°C. Victorian workplaces have introduced measures to mitigate workers' heat exposure risk, mainly focusing on managing risk in extreme heat. In particular, many construction worksites in Victoria have adopted heat thresholds for the cessation of work. Workers at all unionised construction sites cease work when temperatures reach 35°C (CFMEU Victoria, 2015). Meanwhile, Californian workers in outdoor-based industries are entitled to shade, water, first aid and paid cool-down rest periods of 10 minutes per 2 hours when temperatures exceed 35°C (California Division of Occupational Safety and Health, 2006). Texas has not mandated specific heat protections for workers.

The results from the three studies measure the residual risk of heat after accounting for these mitigation policies. Hence, the lesser effect of extreme heat estimated in our study suggests that heat cessation policies introduced in Victoria have positively impacted worker health. Avoiding work as a measure to adapt to extreme temperatures may become less feasible as these temperatures become more common. However, increased frequency of temperature extremes may make other forms of adaption more viable, such as investments in cooling infrastructure and protective clothing designed for use in high temperatures, so changes in the future are difficult to predict.

*5.4 Effect of Past Temperatures*

High temperatures on previous days may cause additional claims, particularly accident types related to reduced concentration or fatigue. Moreover, high night-time temperatures could affect the quality of sleep, which may reduce day-time cognitive performance (Carias et al., 2022). Table 4 presents the estimated effects of yesterday's maximum temperature, the day before yesterday's maximum temperature, and last night's minimum temperature for labourers and tradespersons during 2000–2020.[14] We find that the same-day temperature accounts for the majority of the effect. Conditional on today's temperature, a higher maximum temperature yesterday leads to an increase in claims by 0.14% per 1°C (*p*<0.01) (column 2). The day before yesterday's maximum temperature (column 3) has no effect on claims (-0.01% per 1°C, *p*=0.88), conditional on today's and yesterday's maximum temperatures. Overnight minimum temperature also does not significantly affect the number of claims (-0.04% per 1°C, *p*=0.38), suggesting disturbances to sleep are not driving the main effect (column 4).

---

[14] The correlations between todays' maximum temperature and yesterday's maximum temperature, day before yesterday's maximum temperature and last night minimum temperature are 0.82, 0.71 and 0.66 respectively. Given these correlations, multicollinearity may be a concern in models that include both same-day and past temperatures.

Overall, these results indicate that the primary mechanism driving the relationship between heat and claims operates on the same day as the claim. Combined with previous results, it seems that workers' concentration levels are lower during high temperatures, which causes additional accidents. The 'yesterday' effect indicates that extreme heat may continue to affect workers' concentration even after temperatures have become cooler; however, we expect the types of claims caused by the previous day temperature to be different than those caused by the same-day temperature (e.g. sunburn less likely on the second day).[15] We show the full effect of temperature (today's effect plus yesterday's effect) in the following sections of the paper. This combined effect represents the total effect of a 1°C increase in maximum temperature for a single day, resulting in both same-day and next-day claims.

# 6 Heterogeneity in the Effect of Heat

*6.1 Heterogeneity by Accident Type*

Here we investigate the impact of heat on the nature (e.g. fracture, burn), type (e.g. fall, hit by object) and cause (e.g. machinery, chemicals) of the accident.[16] We again focus exclusively on our sample of high-risk occupations (i.e. labourers and tradespersons) and claims that occurred during years 2000-2020. The estimates in Figure 4 indicate that increases in temperature significantly increase claims for multiple accident natures, with the largest effect occurring for 'burns' (1.57% per 1°C, *p*<0.01), which includes chemical and electrical burns, but does not include sunburn. The smallest effect occurs for 'diseases' (0.19% per 1°C, *p*=0.02), which have approximate affliction dates. This result is consistent with our expectations, since there is no plausible mechanism between short-term fluctuations in weather and disease diagnosis for many disease types (such as cancer).

---

[15] We use the date of affliction for each injury, rather than the claim reporting date. So, we attribute the effect of yesterday's temperature to delayed effects of heat exposure rather than reporting errors.
[16] Occupational health claims relate to a variety of injuries and diseases occurring for a range of hazards. Claims are coded using the Type of Occurrence Classification System (TOOCS) (Australian Safety and Compensation Council, 2008). Each claim is grouped by accident type, accident cause and nature of injury or disease. Classification is structured hierarchically with the first digit of each code representing the major group. This analysis primarily focuses on these major groups to ensure adequate power for estimation.

Unsurprisingly, we find the largest effect for accidents caused by 'heat'. Since 2000, for labourers and tradespersons, a 1°C increase in maximum temperature causes 2.11% more heat-related claims ($p<0.01$). Most of these claims relate to contact with hot objects (62%), while a minority (6%) relate to exposure to environmental heat. This result indicates that heat-related accidents are not primarily due to direct physiological responses to heat, such as heat stroke or dehydration. We also see large and significant effects for accidents seemingly unrelated to heat: hitting objects and being hit by moving objects. This further indicates that the health impacts of heat are broader than heat stress, likely also affecting concentration and psychomotor performance. Laboratory experiments have identified a decrease in task performance when temperatures exceed 24°C (Seppanen et al., 2006). Similarly, heat has been shown to impact decision-making and decrease psychomotor performance (Heyes & Saberian, 2019; Künn et al., 2019). In a survey of Australian workers, 50.1% said they have experienced poor concentration when exposed to heat at work (Humphrys & Newman, 2021). However, we do not find falls significantly increase on hot days (0.09% per 1°C, $p=0.46$), which could also be affected by reduced concentration. This differs from Park et al. (2021), who find a significant increase in falls on hot days.

Mental health-related claims, which may not relate to a single date of affliction, do not significantly increase with temperature. As expected, accidents related to "sound and pressure" are not heightened on hot days since most of these accidents relate to hearing loss from long-term exposure to sound. Similarly, accidents related to "body stressing", which includes injuries and diseases from overuse and repetitive movements, are insensitive to temperature.

We find that accidents caused by all agents are at an increased risk on hot days. Notably, accidents caused by materials/substances and chemicals (e.g. chemicals, toxic substances) occur more in high temperatures. These additional claims could be caused by heat-induced cognitive impairments (Park, 2022). However, an alternate (or perhaps concurrent) explanation for the surplus chemical-related accidents is that heat directly impacts the environment to create additional hazards. For example, heat may increase the volatility of chemicals and create hot surfaces, explosions or fires. To make workers safer, in addition to controls to manage dehydration and heat stress, firms need to manage additional environmental hazards created by heat.

*6.3 Which industries are able to mitigate the risks of heat exposure?*

Table 5 shows the effect of maximum temperature by industry. We present estimates for the ten industries with the highest number of claims for tradespeople and labourers. Within each industry, the effects are generally larger for labourers and tradespersons than for other occupations, which are presented in the second column. Regression outcome means, sample sizes, p-values adjusted for multiple hypothesis testing, and estimates based on same-day temperature only are reported in Appendix H.

Workers in 'Agriculture, Forestry and Fishing' are more negatively affected by heat than workers from other industries. The manual worker estimate for Agriculture, Forestry and Fishing indicates that a 1°C increase in temperature increases the number of claims by 0.76% ($p=0.04$). If we allow for nonlinearity and use temperature bins, rather than the single temperature term, we find for this group of workers 17.8% ($p=0.01$) and 9.9% ($p=0.26$) additional claims when the maximum temperature is 30–33°C and 33–36°C respectively, compared to 18–21°C. The next largest effects are for manual workers in accommodation and food services (0.63 per 1°C; $p=0.587$), health care and social assistance (0.55 per 1°C; $p=0.142$) and transport postal and warehousing (0.53 per 1°C; $p=0.132$) industries, however these effects are not statistically significant at the 5% level. The effect for manual workers in Construction is also large and statistically significant (0.51% per 1°C; $p<0.01$). Workers in the construction and agriculture industries typically spend more time outdoors (National Center for O*NET Development, 2021), suggesting that time outdoors is also an important factor. This is consistent with the findings of Page and Sheppard (2019), Dillender (2021) and Xiang (2014).

*6.4 Characteristics of Workers Vulnerable to Heat Exposure*

Table 6 investigates heterogeneity by sex, age, income and firm size. We consider claims across all industries in the first column, labourers and tradespersons in the second column and labourers and tradespersons in the agriculture and construction industries in the third column. Regression outcome means, sample sizes, p-values adjusted for multiple hypothesis testing and estimates based on same-day temperature only are reported in Appendix I.

Young males have previously been identified as the workers most impacted by heat (Park et al., 2021; Xiang et al., 2014). Indeed, the overall effect of temperature for males is much larger than for females, where effect sizes are 0.35 and 0.03, respectively. Similarly, workers aged 18–35 have a larger effect (0.31% per 1°C) than those aged 36–50 (0.15% per 1°C). However, these results do not consider differences in the gender and age composition of industries and occupations. Labourers and tradespersons in the agriculture and construction industries are disproportionately younger and male. When looking within these high-risk industries and occupations, the point estimate for females is larger than for males (0.81 and 0.48) and is larger for older workers than for middle-aged workers (0.64 and 0.345); though the differences across gender ($p=0.36$) and age ($p=0.42$) are not statistically significant.

To explore heterogeneity by income, we use the pre-injury earnings to identify high, medium and low-wage workers. We find that low-, medium- and high-income earners are similarly impacted by heat, having 0.17%, 0.21% and 0.15% additional claims for each 1°C increase in daily maximum temperature, respectively. This finding is different to Park et al. (2021), who find that lower-income workers are disproportionately affected by heat (the authors use the socioeconomic status of the claimant's postcode to proxy income). When looking at high-risk workers, we find high-income workers have the largest effects. High-income labourers and tradespersons in agriculture and construction have 0.77% more claims per 1°C compared to 0.21% and 0.48% for their low-income and middle-income counterparts.

For all worker and high-risk manual worker samples, we find that workers in large firms are less affected by temperature than those working in medium and small firms. This difference may be a result of heat policies, such as heat thresholds for the cessation of work introduced at large-scale construction sites in Victoria.

## 7  Firm Differences in the Impact of Heat

In this section, we investigate heterogeneity in the effect of heat between different types of firms; specifically, between typically 'safe' (low-claims) and 'unsafe' (high-claims) workplaces. The aim is to inform on the potential impact of firm-level decisions on heat-related risks, which can help us understand whether additional claims are likely to be mitigated through sound risk management practices and policies. Isolating firm-level effects has become a focus of labour economics using linked firm-employee data (Abowd et al., 1999). But this study is rare in investigating the relative importance of firms in managing worker health. A unique feature of our database is that it contains firm identification numbers for each claim. In total, there are claims from 172,615 unique firms during the analysis period.

We use a two-stage approach for this analysis. First, we create a monthly firm-level panel of claims. To identify "high" and "low" claim firms, we use the period 2000-2004. To estimate the effects of temperature, we use the period 2005-2019. To account for (potentially endogenous) openings and closures of firms, we restrict the sample to firms that exist throughout the 20-year sample period. We focus on claims from labourers and tradespersons at firms in the Construction, Manufacturing and Agriculture, Forestry & Fishing industries. We include manufacturing firms in this analysis for a larger sample size and the high proportion of manual workers. These combined restrictions yield a balanced monthly panel of 909 firms over 60 months. We use fixed-effects linear regression to model the monthly ($m$) number of claims for labourers and tradespersons at each firm ($i$). The model contains individual firm fixed-effects ($firm_i$) and time fixed-effects ($\phi_m$) as shown in Equation (2).

$$claims_{im} = firm_i + \phi_m + u_{im} \qquad (2)$$

This regression is estimated separately for nine groups of firms based on the combination of size (small, medium and large) and industry (Construction, Manufacturing and Agriculture, Forestry & Fishing).[17] We use the estimated firm fixed-effect to classify firms as "high claims" (above median) and "low claims" (below median) compared with other firms in the same industry and size.

The second stage of the analysis estimates temperature effects for high- and low-claim firms separately using the main specification used throughout the paper. The results of this analysis are presented in Table 7. We find that the percentage increase in claims is largest at typically 'safer' firms. Low claim firms experience a 0.69% total increase in claims for each additional 1°C in daily maximum temperature, compared to 0.38% per 1°C for high claims firms.[18] Interestingly, the effect at low claims firms is driven by the effect of the same-day temperature (today: 0.71% per 1°C; yesterday: -0.02% per 1°C), while the effect at high claim firms is primarily driven by the previous day's temperature (today: 0.07% per 1°C; yesterday 0.31% per 1°C).

These results indicate that the adverse impact of heat is not limited to typically 'unsafe' firms. One possible explanation is that heat-related accidents are challenging to prevent, even for 'safe firms' that have invested in worker safety. Another explanation is that safety risks caused by heat are uncorrelated with or maybe even negatively correlated with other safety risks. For instance, a survey of health and safety representatives identified wearing personal protective equipment as a risk factor for heat-related injuries, as certain clothing types can lead to higher body temperatures (Varghese et al., 2020). A third possible explanation is that low-claim firms are not actually safe, but instead have low numbers of claims because they discourage reporting of incidents. The higher estimate for such firms may arise because heat-related injuries are difficult to avoid reporting.

---

[17] Employer size based on the remuneration in 2010/11 deflated to 2005/06 dollars (<1Million = Small, 1-20 Million = Medium, >20 Million = Large). Primary industry is determined as the industry of most claims.
[18] We also estimate effects using temperature bins and find the shape of the temperature-claims is similar for both groups and to the overall sample presented in Figure 3.

## 8 Costs of Heat on Workers' Health

We now estimate the costs associated with heat-related accidents. The total cost of occupational health is substantial, with work-related injuries in Victoria estimated to cost AU$14.6billion in the 2012/13 financial year, which was equivalent to 4.4% of gross state product (Safe Work Australia, 2015). This estimate is inclusive of both compensated and non-compensated injuries and diseases. For our calculations, we use unit costs of AU$258,700 for claims and AU$126,500 for all occupational injuries and diseases (Safe Work Australia, 2015).[19] These costs include the human-related costs for injuries and illnesses occurring at work, but not costs such as damage to property. Wellbeing losses, such as pain and suffering, are also not included. The largest contribution to the unit cost is human capital costs that reduce long-term earnings and productivity, which are borne by the workers and society, rather than firms. Therefore, any increased costs from warming are likely to be predominately borne by workers through loss of future earnings. SafeWork Australia (2015) estimate that 74% of the costs are borne by workers, 21% by the community and only 5% by firms.

The Earth's climate has warmed over recent decades, with the seven years since 2014 being the hottest years on record (GISTEMP Team, 2021). Even with a considerable reduction in greenhouse gas emissions, global average temperatures will likely be 1.5°C above pre-industrial levels by 2040 (Masson-Delmotte et al., 2021). A simple calculation using the linear effect of temperature on worker health implies that a warming of Victoria's climate by 1.5°C would result in an additional 0.38% occupational health claims for all workers, assuming current adaptation measures are maintained.[20] This equates to an additional 156 claims per annum at a cost of AU$40.3 million for Victoria. If we consider all injuries and diseases, regardless of compensation, the annual cost of

---

[19] Safe Work Australia estimate there were 115,415 total occupational injuries and diseases for FY12/13 in Victoria, while for the same period there were 41,586 claims submitted to WorkSafe Victoria recorded in the CRD. To estimate costs inclusive of non-compensated claims, we make a simplifying assumption that the injuries and diseases without compensation also increase with temperature at the same rate. We use the number of cases from FY12/13 as claims can be submitted over several years following the affliction.
[20] Recall, we find that an additional 0.42% occupational health claims per 1°C for labourers and tradespersons (Table 4, column 2). The full effect of temperature (today and yesterday) for all workers is 0.25% per 1°C for all workers. Here, we assume that the maximum temperature is increased by 1.5°C every day and does not account for non-linearities in effects. Adaptation to heat, such as avoidance to work on hot days, may change in the future, but is difficult to predict. Our calculations assume that current adaptation measures are maintained.

1.5°C of warming is AU$54.8 million, of which labourers and tradespersons account for AU$37.4 million.

Victoria's climate has experienced considerable warming throughout the 35-year analysis period. We next calculate the approximate cost of the observed warming during this period. The average daily maximum temperature in Victoria typically exceeded 30°C on 25 days each year in the 1980s. To get a single value for daily maximum temperature in Victoria we take the average maximum temperature of Victorian postcodes weighted by the area's total claims. Since 2015, Victoria now averages 36 days each year over 30°C. Similarly, there has been a reduction in the number of days with low maximum temperatures. We determine the annual change in the number of days per year occurring in each 3°C temperature bin. Then, we calculate the corresponding change in the annual number of claims using estimates for these temperature bins. Based on these calculations, we estimate that past warming has an average per annum cost of AU$26.3 million for claims and AU$35.6 million for all workplace injuries and diseases in Victoria (detailed calculations are provided in Appendix K). For labourers and tradespersons, the annual costs are AU$15.6 million for claims and AU$21.2 million overall.

## 9 Conclusion

The paper has four key findings. First, high outside temperature has a significant negative effect on workers' health, particularly for manual workers in outdoor-based industries. This effect is costly, with a 1.5°C increase in temperatures costing an additional AU$54.8million each year in heat-related injuries and diseases at work. Second, although we find larger effects for younger and male workers overall, by estimating effects for subgroups within high-risk industries and occupations, we show that these differences can be attributed to differences in the gender and age composition of industries and occupations. Third, our unique firm-level panel reveals that even firms with a low number of claims overall are sensitive to temperature, which may suggest that preventative measures to make workplaces safer is difficult. Finally, adverse temperature effects are observed throughout the 35-year analysis period, despite policy attempts to address the risks. Notably, the effects are largest in recent years, which have also been the hottest, supporting the proposal that there are limits on workers' ability to adapt.

Our main finding of increased risk on hot days aligns with results from the U.S. in both magnitude and the near-linear shape of the temperature-claims relationship (Dillender, 2021; Page & Sheppard, 2019; Park et al., 2021). This demonstrates that the negative effects of temperature on workers' health apply despite differences in climate, occupational health systems, and protective measures. Consistent with previous studies, we also find that manual workers in outdoor-based industries are most sensitive to the impacts of heat exposure. Importantly, these occupations have fewer viable options for adaptation, such as air conditioning or the ability to work from home, and may become increasingly vulnerable as hot days occur more often. We show the largest heat effects for high-risk occupations occurred in recent years, which have also been the hottest. Although the heat effect based on all occupations was larger prior to 2000, the effect has not reduced when looking within occupations. This suggests that the decrease is likely driven by changes in the occupational composition away from high-risk occupations rather than effective adaptation within occupations, and highlights the importance of accounting for changes in the labour force over time in the climate-adaptation literature. We also find that heat causes multiple types of accidents, which aligns with the results from Park et al. (2021). This reinforces the importance of reduced concentration as a likely mechanism.

The underlying presumption in the climate-economy literature is that the observed effects of heat are a result of human physiological limits (Heal & Park, 2016; Park et al., 2021). However, our results suggest that the mechanisms producing the adverse impacts of heat on workers' health may extend beyond these physiological limits. High temperatures may create a more dangerous environment with additional hazards such as hot surfaces, chemicals and fires. The direct effect of heat on the environment is an important insight for the broader literature on the effects of heat on economic and health outcomes. For instance, reported excess mortality (Barreca et al., 2016; Longden, 2018) and reduced worker productivity (Zhang et al., 2018) on hot days may also plausibly result from factors unrelated to human sensitivity to climate (e.g. machinery malfunctioning). Effective heat policy should not only address workers' physiological needs but also the hazards around them.

**Tables and Figures**

**Table 1. Summary of Claims in the Compensation Research Database**

|  | All Occupations | | | Labourers and Tradespersons | | |
|---|---|---|---|---|---|---|
|  | *All Claims* | *Sep1985-Dec1999* | *Jan2000-Jul2020* | *All Claims* | *Sep1985-Dec1999* | *Jan2000-Jul2020* |
| *Demographic* | | | | | | |
| Male, % | 70 | 73 | 67 | 83 | 83 | 82 |
| Age 18–35, % | 41 | 46 | 35 | 45 | 47 | 41 |
| Age 36–50, % | 36 | 35 | 37 | 33 | 32 | 34 |
| Age 51–64, % | 21 | 17 | 25 | 19 | 18 | 22 |
| *Industry* | | | | | | |
| Agriculture, Forestry & Fishing, % | 2 | 2 | 2 | 3 | 2 | 3 |
| Construction, % | 9 | 9 | 10 | 16 | 14 | 19 |
| Manufacturing, % | 26 | 30 | 21 | 39 | 42 | 34 |
| Other Industry, % | 63 | 60 | 67 | 43 | 43 | 44 |
| N | 2,156,960 | 1,130,864 | 1,026,096 | 1,062,518 | 644,353 | 418,165 |

**Note:** Based on claims in the Compensation Research Database (CRD) for afflictions occurring between 2nd September 1985 and 31st July 2020. Figures are a percentage of claims corresponding to the descriptor in column 1.

**Table 2. Estimated Effect of Maximum Temperature on Occupational Health Claims**

|  | (1) Main Spec. | (2) PM2.5 | (3) Calendar Date | (4) Separate F.E.s |
|---|---|---|---|---|
| *Coefficient* | 0.24 | 0.20 | 0.27 | 0.22 |
| *(Semi-Elasticity %)* | (0.020) | (0.021) | (0.020) | (0.021) |
| Month-Year-Postcode FE | ✓ | ✓ | ✓ | ✗ |
| Air Pollution (PM2.5) | ✗ | ✓ | ✗ | ✗ |
| Calendar Date FE | ✗ | ✗ | ✓ | ✗ |
| Postcode-Month FE | ✗ | ✗ | ✗ | ✓ |
| Month-Year FE | ✗ | ✗ | ✗ | ✓ |
| N | 3,533,474 | 3,533,474 | 3,533,474 | 3,533,474 |

**Notes:** Estimated effects of daily maximum temperature (°C) on the number of claims in a postcode as outcome variable using Poisson regression. Semi-elasticity represents a percentage increase or decrease on the number of claims. Column (1) is the main specification with month-year-postcode fixed-effects. Columns (2) and (3) adds to the main specification with the additional controls for air pollution and calendar date fixed-effects respectively. Column (4) uses separate month-year and postcode-month fixed effects. All regressions include day of week fixed effects and precipitation (mm) controls. Robust standard errors, clustered by postcode, in parentheses below estimates. Outcome mean is 0.54 claims per postcode-day.

**Table 3. Maximum Temperature Estimates by Occupation Over Time, Semi-elasticity (%)**

|  | **All Years** | **1985–1999** | **2000–2020** |
|---|---|---|---|
| All Occupations | 0.24 | 0.27 | 0.20 |
|  | (0.020) | (0.027) | (0.026) |
| High Risk Occupations | 0.35 | 0.36 | 0.34 |
|  | (0.027) | (0.035) | (0.038) |
| *- Tradespersons and related workers* | 0.38 | 0.38 | 0.36 |
|  | (0.042) | (0.051) | (0.057) |
| *- Labourers and related workers* | 0.33 | 0.34 | 0.31 |
|  | (0.033) | (0.042) | (0.053) |
| Other Occupations | 0.12 | 0.13 | 0.10 |
|  | (0.028) | (0.044) | (0.035) |
| *- Production and transport workers* | 0.16 | 0.16 | 0.16 |
|  | (0.045) | (0.066) | (0.062) |
| *- Intermediate clerical, sales and service workers* | 0.15 | 0.19 | 0.13 |
|  | (0.070) | (0.146) | (0.085) |
| *- Associate professionals* | 0.13 | 0.07 | 0.16 |
|  | (0.071) | (0.112) | (0.094) |
| *- Professionals* | 0.08 | 0.10 | 0.06 |
|  | (0.057) | (0.090) | (0.081) |
| *- Elementary clerical, sales and service workers* | 0.06 | 0.11 | -0.04 |
|  | (0.101) | (0.132) | (0.132) |
| *- Managers and administrators* | 0.06 | 0.25 | -0.11 |
|  | (0.110) | (0.156) | (0.144) |
| *- Advanced clerical and service workers* | 0.04 | -0.07 | 0.13 |
|  | (0.179) | (0.215) | (0.278) |

**Notes:** Estimated effects of daily maximum temperature (°C) on the number of claims in a postcode as outcome variable. All regressions use Poisson Regression Month-Year-Postcode fixed-effects as per column (1) of Table 2. Semi-elasticity represents a percentage increase or decrease on the number of claims. Claims categorised using ASCO occupation codes. "Other" combines all claims for clerical, sales, service, production, transport, professional, associate professional, managers and administrators and workers into a single group. Robust standard errors, clustered by postcode, parenthesised below estimates. Outcome mean and sample sizes are provided in Appendix F1.

**Table 4. Effect of Past Temperature on Occupational Health Claims**

|  | (1) | (2) | (3) | (4) |
|---|---|---|---|---|
| Today Maximum Temperature | 0.34 (0.04) | 0.28 (0.04) | 0.28 (0.04) | 0.28 (0.04) |
| Yesterday Maximum Temperature | - | 0.14 (0.05) | 0.14 (0.05) | 0.16 (0.05) |
| Day Before Yesterday Maximum Temperature | - | - | -0.01 (0.05) | - |
| Last Night Minimum Temperature | - | - | - | -0.04 (0.08) |
| Combined | 0.34 (0.04) | 0.42 (0.05) | 0.41 (0.05) | 0.39 (0.06) |

**Notes:** Estimated effects of daily maximum temperature (°C) for labourers and tradespersons 2000–2020. All regressions use Poisson Regression Month-Year-Postcode fixed-effects as per column (1) of Table 2. Outcome variable is the number of claims in a postcode as outcome variable. Robust standard errors clustered by postcode. Semi-elasticity represents a percentage increase or decrease on the number of claims. Minimum temperature in 24 hours before 9am (local time). Outcome mean is 0.24 claims per postcode-day (N=1,603,804).

**Table 5. Maximum Temperature Estimates by Industry, Semi-elasticity (%)**

| Industry | Labourers and Tradespersons | Other Occupations |
|---|---|---|
| Agriculture, Forestry and Fishing | 0.76 (0.269) | 0.11 (0.429) |
| Accommodation and Food Services | 0.63 (0.348) | -0.04 (0.218) |
| Health Care and Social Assistance | 0.55 (0.245) | 0.07 (0.096) |
| Transport, Postal and Warehousing | 0.53 (0.227) | 0.43 (0.120) |
| Construction | 0.51 (0.114) | 0.38 (0.222) |
| Manufacturing | 0.36 (0.079) | 0.10 (0.107) |
| Administrative and Support Services | 0.35 (0.209) | -0.16 (0.248) |
| Wholesale Trade | 0.17 (0.207) | 0.28 (0.158) |
| Retail Trade | -0.10 (0.234) | 0.11 (0.149) |
| **All industries** | 0.42 (0.045) | 0.13 (0.039) |

Notes. Estimates for the combined effect of today's and yesterday's maximum temperature on number of claims using Poisson regression. Estimates displayed as a semi-elasticity, as a percentage increase or decrease in the number of claims. All regressions include Month-Year-Postcode fixed-effects as per column (1) of Table 2. Robust standard errors clustered at postcode level in parenthesis. 2000–2020. Industries with the most claims for labourers and tradespersons are displayed. See Appendix H for regression sample sizes, outcome means and adjusted p-values.

**Table 6. Maximum Temperature Estimates by Worker Characteristic, Semi-elasticity (%)**

|  | All Workers | Labourers and Tradespersons | Labourers and Tradespersons in Agriculture or Construction |
|---|---|---|---|
| <u>Sex</u> | | | |
| Male | 0.35 (0.038) | 0.46 (0.053) | 0.48 (0.090) |
| Female | 0.03 (0.056) | 0.19 (0.119) | 0.81 (0.346) |
| | | | |
| <u>Age</u> | | | |
| Age 18–35, % | 0.31 (0.046) | 0.42 (0.070) | 0.48 (0.130) |
| Age 36–50, % | 0.15 (0.053) | 0.36 (0.088) | 0.45 (0.157) |
| Age 51–64, % | 0.31 (0.058) | 0.49 (0.102) | 0.64 (0.182) |
| | | | |
| <u>Earnings</u> | | | |
| Low earnings | 0.17 (0.076) | 0.23 (0.120) | 0.21 (0.182) |
| Middle earnings | 0.21 (0.070) | 0.25 (0.108) | 0.48 (0.187) |
| High earnings | 0.15 (0.083) | 0.31 (0.104) | 0.77 (0.216) |
| | | | |
| <u>Firm Size</u> | | | |
| Small | 0.34 (0.070) | 0.43 (0.098) | 0.40 (0.134) |
| Medium | 0.30 (0.045) | 0.41 (0.070) | 0.63 (0.144) |
| Large | 0.15 (0.058) | 0.27 (0.095) | 0.23 (0.208) |
| **All Claims** | 0.25 (0.030) | 0.42 (0.045) | 0.55 (0.106) |

**Notes:** Estimates for combined effect of today's and yesterday's maximum temperature on number of claims using Poisson regression for claims with afflictions occurring between 2000 and 2020. Estimates displayed as a semi-elasticity, as a percentage increase or decrease in the number of claims. All regressions include Month-Year-Postcode fixed-effects as per column (1) of Table 2. Low, medium and high wage based on terciles for the year and population of the respective column where provided. Robust standard errors clustered at postcode level in parenthesis. Firm size is based on total annual remuneration in 2010/11 deflated to 2005/06 dollars (<1Million = Small, 1–20 Million = Medium, >20 Million = Large). See Appendix I for regression sample sizes, outcome means and adjusted p-values.

**Table 7 Effect of Maximum Temperature on Occupational Health Claims by Firm**

|  | "Low Claims" Firm | "High Claims" Firm |
|---|---|---|
| Today's Temperature | 0.71 (0.302) | 0.07 (0.171) |
| Yesterday's Temperature | -0.02 (0.427) | 0.31 (0.156) |
| Combined | 0.69 (0.343) | 0.38 (0.165) |
| Number of firms | 454 | 455 |
| Outcome mean | 0.07 | 0.11 |
| N (postcode-days) | 111,514 | 296,694 |

**Notes:** Estimates for combined effect of today's and yesterday's maximum temperature on number of claims using Poisson regression. Based on claims for labourers and tradespersons occurring between 2005–2019 for a subset of firms in Construction, Manufacturing and Agriculture, Forestry & Fishing. Firms are classified as "high claims" and "low claims" firms compared with other firms in the same industry and size using the individual firm fixed-effect (firms above the median considered "high claims"). Regressions include Month-Year-Postcode fixed-effects as per column (1) of Table 2. Robust standard errors clustered at postcode level in parenthesis.

**Figure 1. Number of claims in the Compensation Research Database by month**

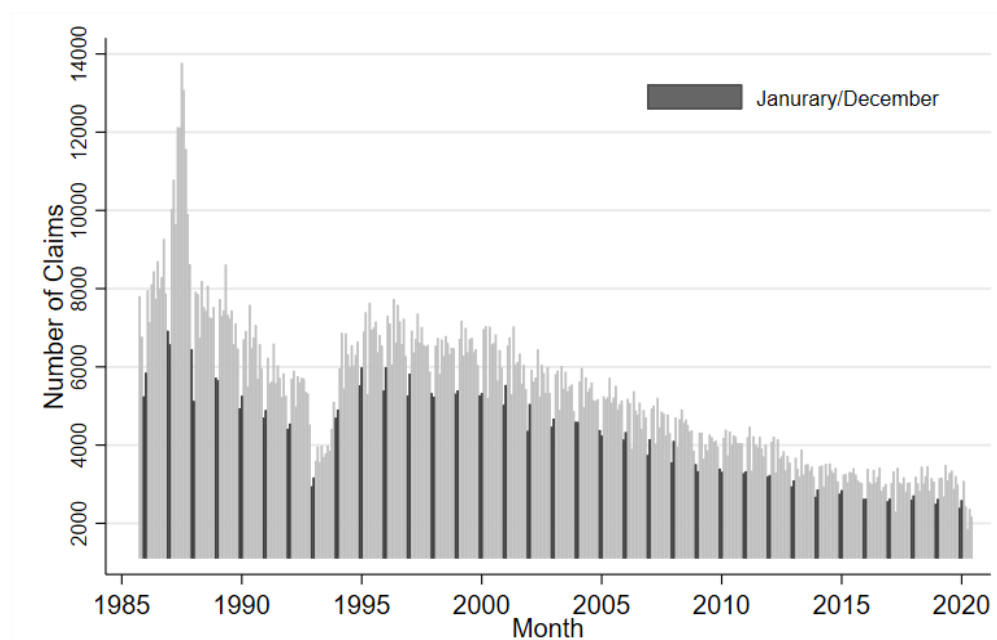

**Notes:** Number of claims in the Compensation Research Database (CRD) by month based on date of affliction. The CRD records all claims submitted to Victoria's mandatory workers' compensation insurance scheme since its inception in 1985.

**Figure 2. Effect of Maximum Temperature on Occupational Health Claims over Time**

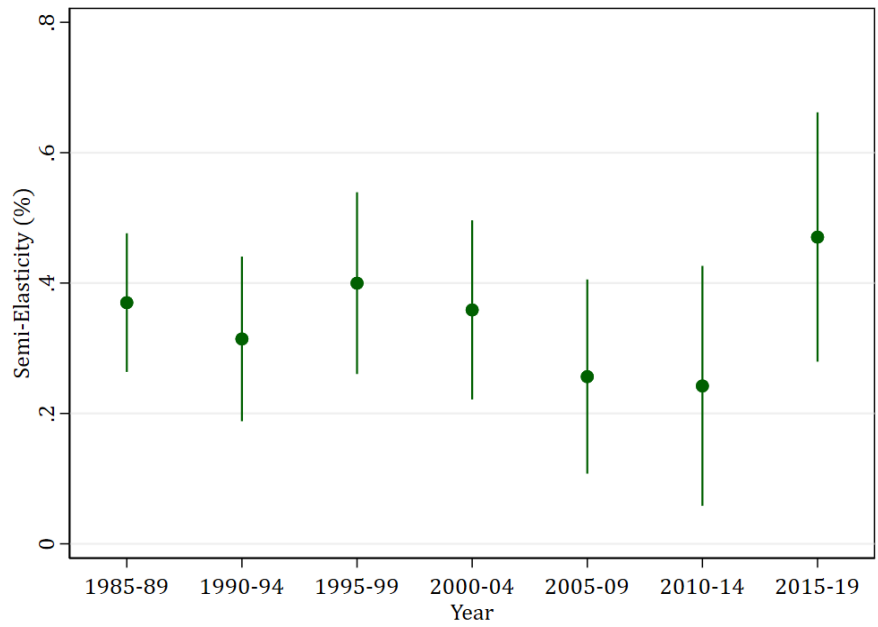

**Notes:** Estimated effects of daily maximum temperature (°C) for Labourers, Tradespersons and Related workers using Poisson regression with Month-Year-Postcode fixed-effects as per column (1) of Table 2. Outcome variable is the number of claims in a postcode as outcome variable. Robust standard errors clustered by postcode. Semi-elasticity represents a percentage increase or decrease on the number of claims. 95% confidence intervals displayed.

**Figure 3. Effect of Maximum Temperature on Occupational Health Claims**

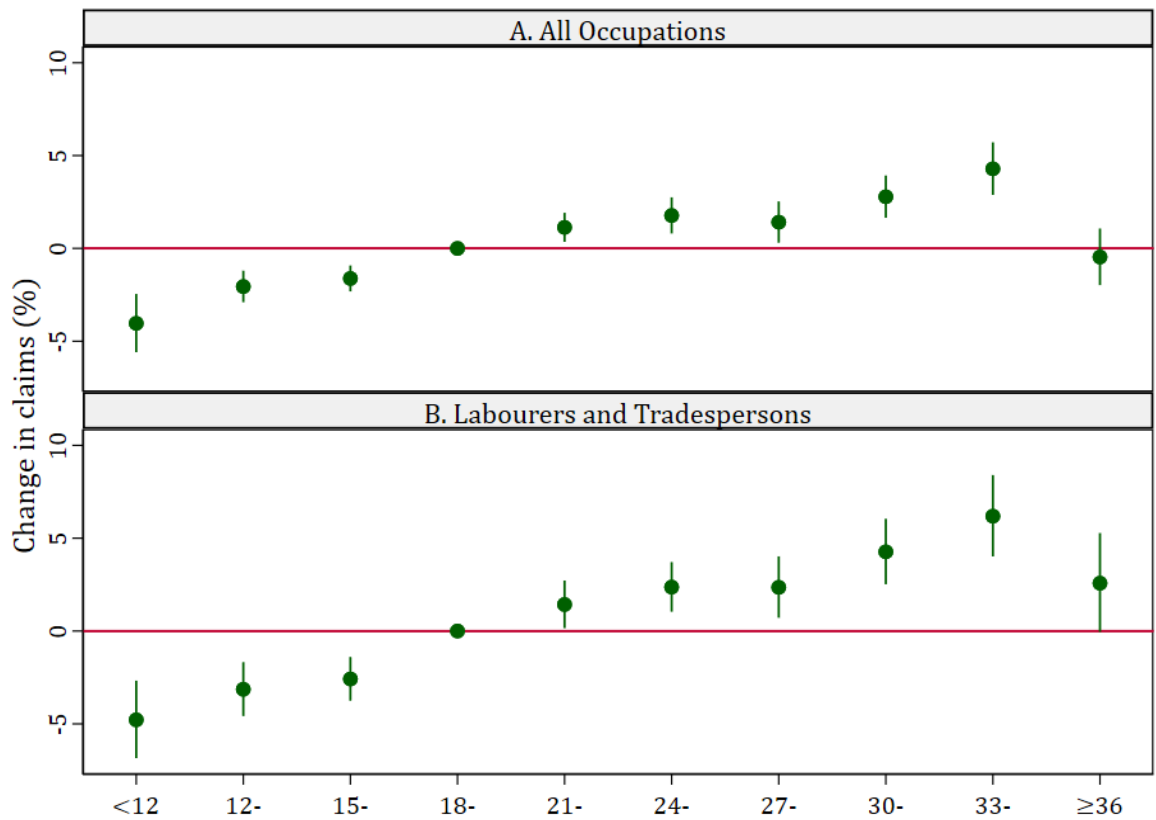

**Notes:** Estimated effects of daily maximum temperature (°C) on occupational health claims using Poisson regression with month-year-postcode fixed-effects as per column (1) of Table 2. Panel A. shows all occupations and Panel B. shows labourers, tradespersons and related workers. Effects shown for the period 2000–2020. Outcome variable is the number of claims in a postcode as outcome variable. Robust standard errors clustered by postcode. Semi-elasticity represents a percentage increase or decrease on the number of claims. 95% confidence intervals displayed.

**Figure 4. Coefficient Estimates by Accident Nature, Type and Cause**

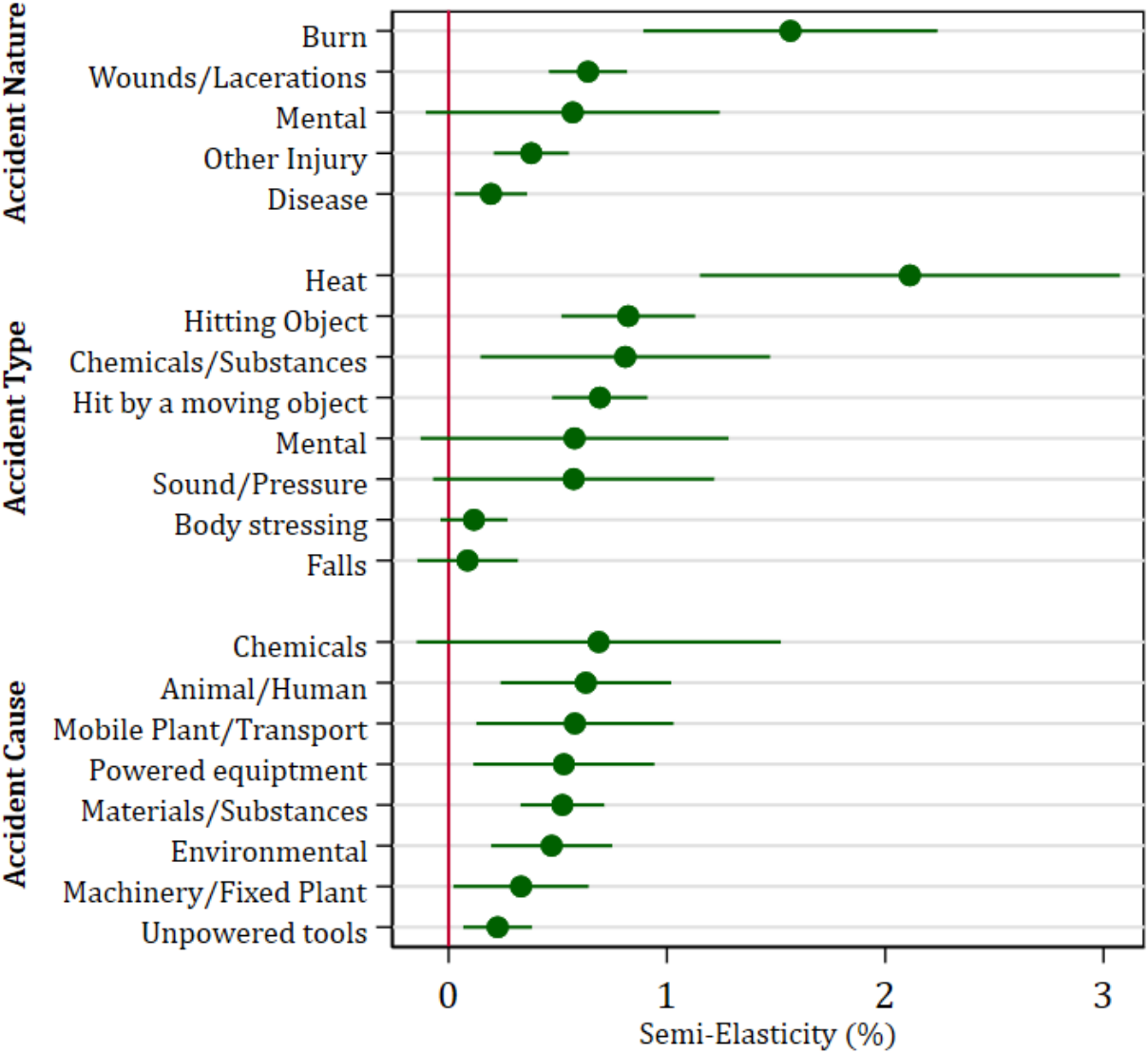

**Notes:** Estimates for the combined effect of today's and yesterday's maximum temperature on number of claims. All regressions use Poisson Regression Month-Year-Postcode fixed-effects as per column (1) of Table 2. Classification of accident nature, type and cause of affliction are based on TOOCS. Effects shown for the period 2000–2020 in high-risk occupations (labourers and tradespersons). Standard errors clustered by postcode, 95% confidence interval displayed.

**Appendices**

Appendix A. Summary of Changes to Workers Compensation in Victoria

A more extensive list of changes can be found in Safe Work Australia's "*Comparison of Workers Compensation Arrangements in Australia and New Zealand*" (2019). For a detail history of workers' compensation in Victoria, refer to Stylianou (2011). This list draws upon dates listed in these two sources.

**1985** Inception of the "WorkCare" scheme on 1st September 1985 under the Accident Compensation Act 1985.

**1987** Amendment to the Accident Compensation Act to address financial viability of the scheme reducing access to benefits and increasing the grounds to suspended or terminate benefits.

**1992** Restricting weekly benefits for workers with a partial work capacity, establishing expert Medical Panels to determine medical questions, limited access to common law (seriously injured workers), and reinstating the right to sue for economic loss.

**1993** WorkCover Insurance Act 1993 made Victorian WorkCover Authority to manage the scheme. Employer premiums based on their safety track record rather than industry-risk.

**1997** Significantly changing the structure of weekly benefits, death benefits and impairment benefits. Removal of ability to make common law claims, making the scheme entirely "no-fault".

**2006** Extended weekly benefits entitlement period from 104 to 130 weeks, increased payments for workers with a partial work capacity and increasing death benefits.

**2010** The Accident Compensation Amendment Act 2010 increased many of the benefits workers are entitled to, including an almost a doubling of lump sum death benefits, increased benefits for workers suffering permanent impairment, a five-fold increase in benefits awarded to workers who suffer a serious psychiatric impairment and increased weekly payments, particularly for long-term injured workers

Appendix B. Distribution of Daily Claims

**Table B1. Distribution of Daily Occupational Health Claims per Postcode in Victoria**

| Number of Claims | Number of Postcode-Days | Percent of Postcode-Days |
|---|---:|---:|
| 0 claims | 2,509,330 | 71.0 |
| 1 claim | 628,902 | 17.8 |
| 2 claims | 205,981 | 5.8 |
| 3 claims | 87,397 | 2.5 |
| 4 claims | 41,943 | 1.2 |
| 5 claims | 22,373 | 0.6 |
| 6 or more claims | 37,548 | 1.1 |

**Notes:** Distribution of the number of claims in each postcode in the Compensation Research database based on date of affliction. 1985–2020.

Appendix C. Cross Validation of Spatial Interpolation Methods

There are many spatial interpolation methods available for researchers to estimate the value of maximum temperature at a point of interest using observations from ground stations. Predominately, studies in the climate-economy literature simply take the value of the closest station (Park et al., 2020; Xiang et al., 2014), or a weighted average based on distance from the station (Dillender, 2021). We test the performance of four different methods to spatially interpolate daily maximum temperature from ground station observations using cross-validation. We find that more methods that include elevation and distance to coast have the lowest root mean square error (RMSE) in the context of Victoria.

The cross-validation compares the known values of temperature at each station with estimated values using all other stations. We estimated the maximum temperature using each method for each weather station in Victoria using observations from all other stations[21]. The estimated values were compared with the actual observed readings at the station to calculate the mean error (M.E.), mean absolute error (MAE) and root mean square error (RMSE). The cross validation was calculated using the 1st, 10th and 20th days of each month from August 1985 to August 2020. This gave variation across months and years for the analysis period.

The simplest method is to take the value from the closest station. The second method, inverse distance squared (IDS), takes a weighted average of the five closeted observations, with closer stations having higher weighting. Next, we use Ordinary Least Squares Regression (OLS), which includes latitude, longitude, elevation and distance to coast. Finally, we use IDS to interpolate the residuals from the OLS regression, "Residual Inverse Distance Squared (RIDS). The rationale for these residual methods is that OLS models the deterministic portion of the temperature, while the weighted averages of residuals account for local stochasticity.

The RMSE of each method is shown below in Figure C1 and Figure C2. They show that the method based on residuals from OLS regression have the best performance throughout the entire 35-year analysis period. This shows that in Victoria, it is important to consider elevation and distance to coast when measuring maximum temperature. Just taking the value from the closest station has the highest measurement error.

Table C1. Shows the impact of using different interpolation methods on our main results. We see only a small amount of attenuation of estimates using less accurate methods. This is likely due to the majority of claims being in metropolitan Melbourne, where there is a higher density of stations.

**Table C1. Maximum Temperature Estimates Using Different Spatial Interpolation Methods**

| Interpolation Method | All Occupations | Labourers and Tradespersons |
|---|---|---|
| Residual Inverse Distance Squared | 0.24 (0.020) | 0.35 (0.027) |
| Inverse Distance Squared | 0.24 (0.020) | 0.35 (0.027) |
| Closest Station | 0.23 (0.019) | 0.34 (0.027) |

**Notes:** Effects of maximum temperature on number of occupational health claims using Poisson regression using three different measurements for temperature – Residual Inverse Distance Squared, Inverse Distance Square and Closest Station and. 1985–2020.

---

[21] We use values from all stations in Victoria and the bordering states of New South Wales and South Australia.

**Figure C1. Comparison of Spatial Interpolation Methods by Year**

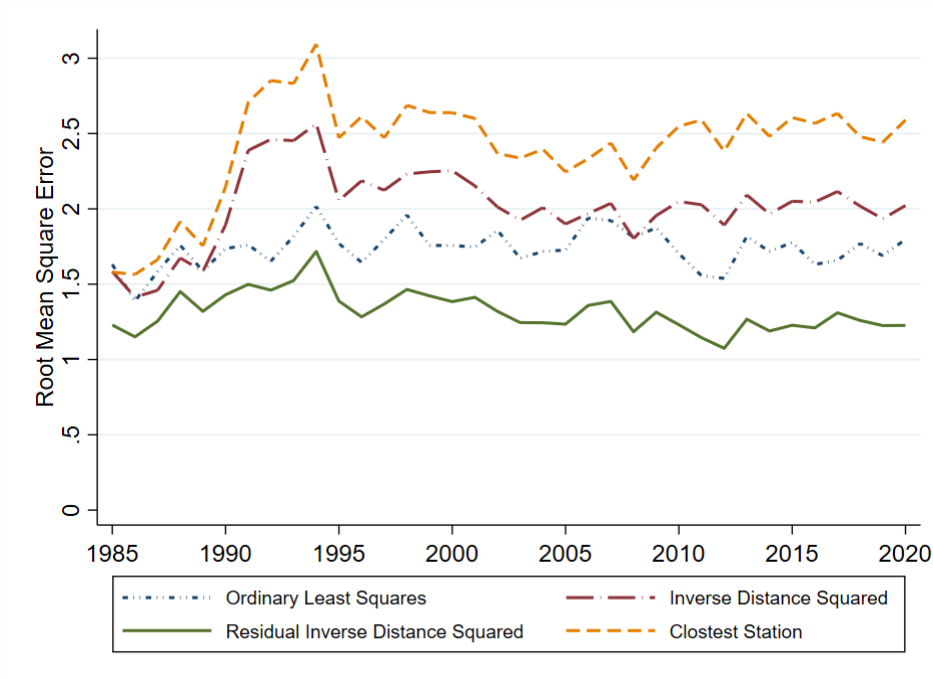

**Notes:** Root mean square error (RMSE) of maximum temperature between actual ground station observations and interpolated values using observations from other ground stations. Based on all weather stations in Victoria recording with temperature observations from 1985–2020.

**Figure C2. Comparison of Spatial Interpolation Methods by Month**

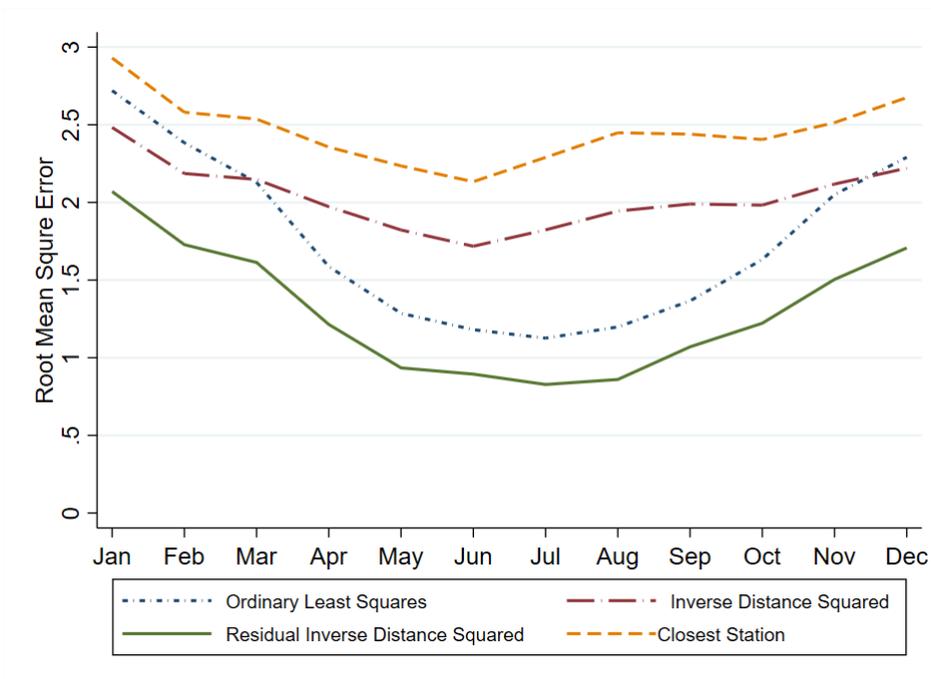

**Notes:** Root mean square error (RMSE) of maximum temperature between actual ground station observations and interpolated values using observations from other ground stations. Based on all weather stations in Victoria recording with temperature observations from 1985–2020.

Appendix D. Climate of Victoria and Weather Stations

**Figure D1. Number of days over 30°C and 35°C in Melbourne by year**

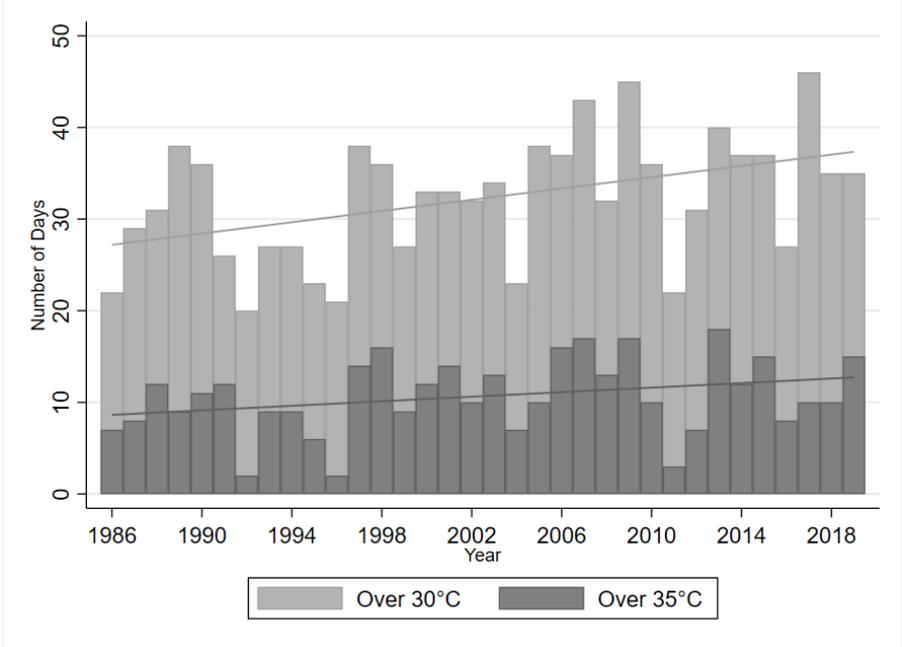

**Notes:** Based on daily maximum temperatures in Melbourne (postcode 3000) from September 1986 to July 2020.

Appendix E. Robustness of Results

**Table E1. Effect of Maximum Temperature on Occupational Health Claims, Robustness**

|  | (1) | (2) | (3) | (4) | (5) |
|---|---|---|---|---|---|
|  | Main Spec. | Dry Days | Including Weekends | Accepted Claims | OLS |
| Coefficient | 0.24 | 0.23 | 0.19 | 0.22 | 0.24 |
| (Semi-Elasticity %) | (0.020) | (0.025) | (0.018) | (0.022) | (0.031) |
| Month-Year-Postcode FE | ✓ | ✓ | ✓ | ✓ | ✓ |
| Air Pollution (PM2.5) | ✗ | ✗ | ✗ | ✗ | ✗ |
| Calendar Date FE | ✗ | ✗ | ✗ | ✗ | ✗ |
| Postcode-Month FE | ✗ | ✗ | ✗ | ✗ | ✗ |
| Month-Year FE | ✗ | ✗ | ✗ | ✗ | ✗ |
| Outcome Mean | 0.54 | 0.61 | 0.42 | 0.46 | 0.31 |
| N | 3,533,474 | 1,631,266 | 5,083,106 | 3,341,613 | 6,331,450 |

**Notes:** Estimated effects of daily maximum temperature (°C) on the number of claims in a postcode as outcome variable. All regressions include day of week fixed effects and precipitation (mm) controls. Columns (1) to (4) use Poisson regression. Semi-elasticity represents a percentage increase or decrease on the number of claims. Column (1) is the main specification with month-year-postcode fixed-effects. Columns (2) restricts the sample to only dry days and column (3) includes weekend days. Column (4) restricts the outcome only accepted claims. effects. Column (5) is estimated by OLS, displayed estimate has been transformed for comparable interpretation as a semi-elasticity (%) dividing by the outcome mean. Robust standard errors, clustered by postcode, in parentheses below estimates.

**Figure E1. Maximum Temperature Estimates by Claim Status**

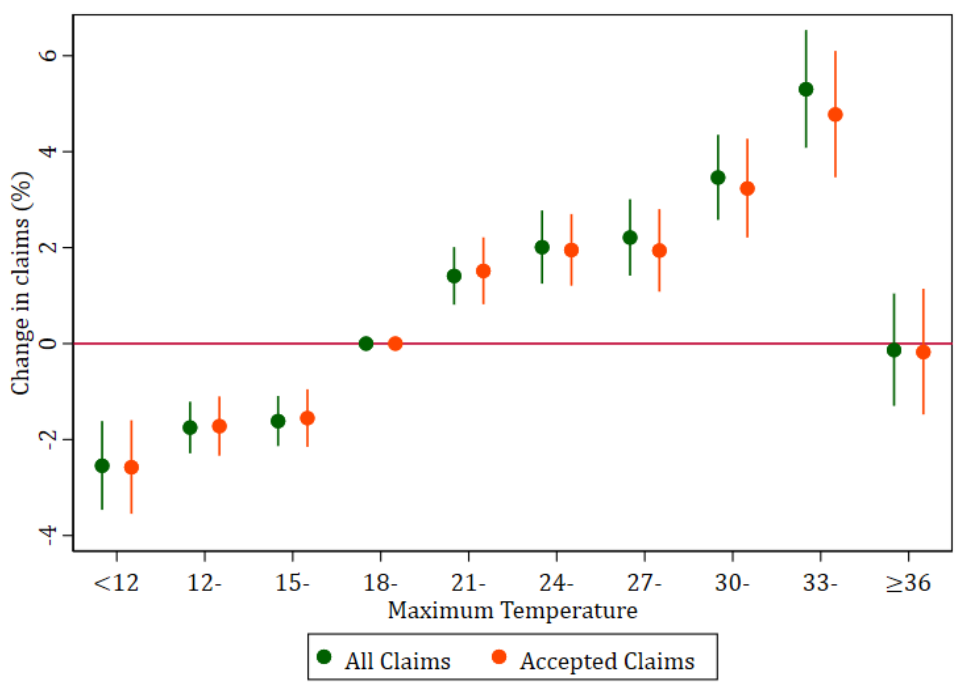

**Notes:** Estimated effects of daily maximum temperature (°C) for all occupations 1985–2020 using Poisson regression with robust standard errors clustered by postcode. Outcome variable is the number of claims in a postcode as outcome variable. semi-elasticity represents a percentage increase or decrease on the number of claims. 95% confidence intervals displayed.

**Figure E2. Maximum Temperature Estimates by Regression Type**

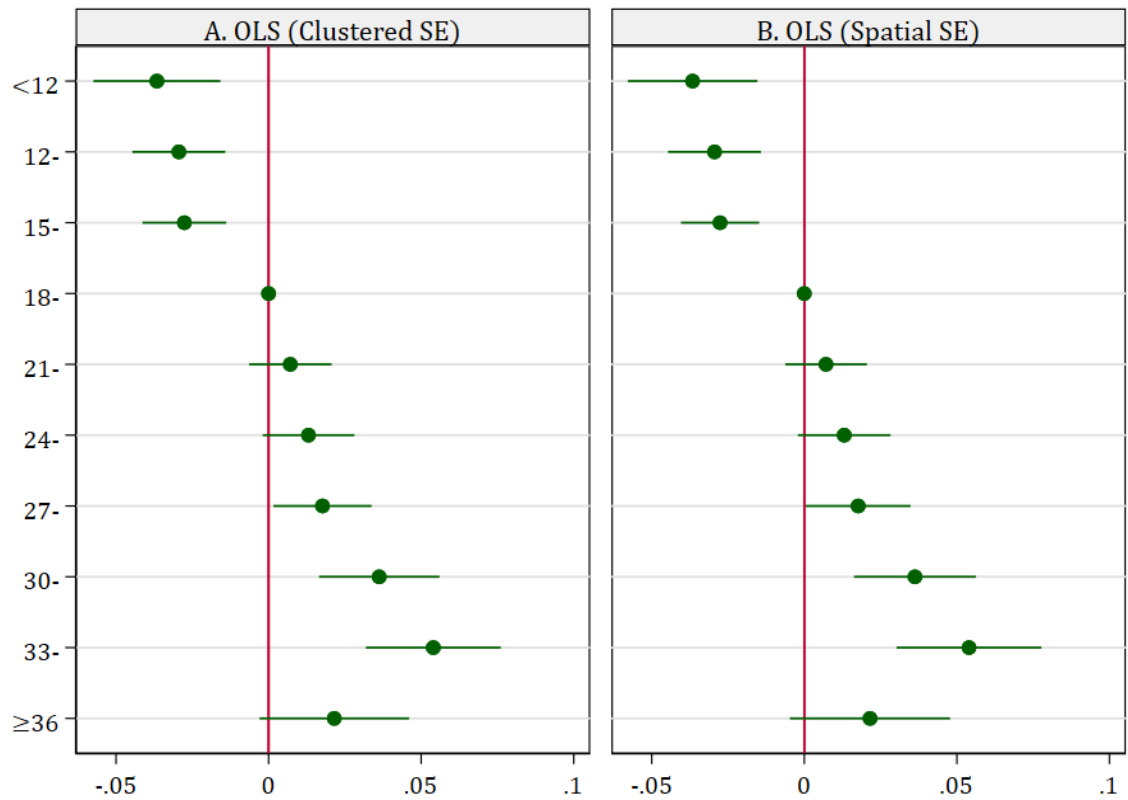

**Notes:** Estimates based on all occupations 2000–2020. OLS regression uses the number of claims divided by the average number of claims in the year as the outcome variable. Panel A shows standard errors clustered at the postcode level and Panel B uses Conley spatial standard errors.

**Figure E3. Maximum Temperature Estimates on Dry Days vs. All Days**

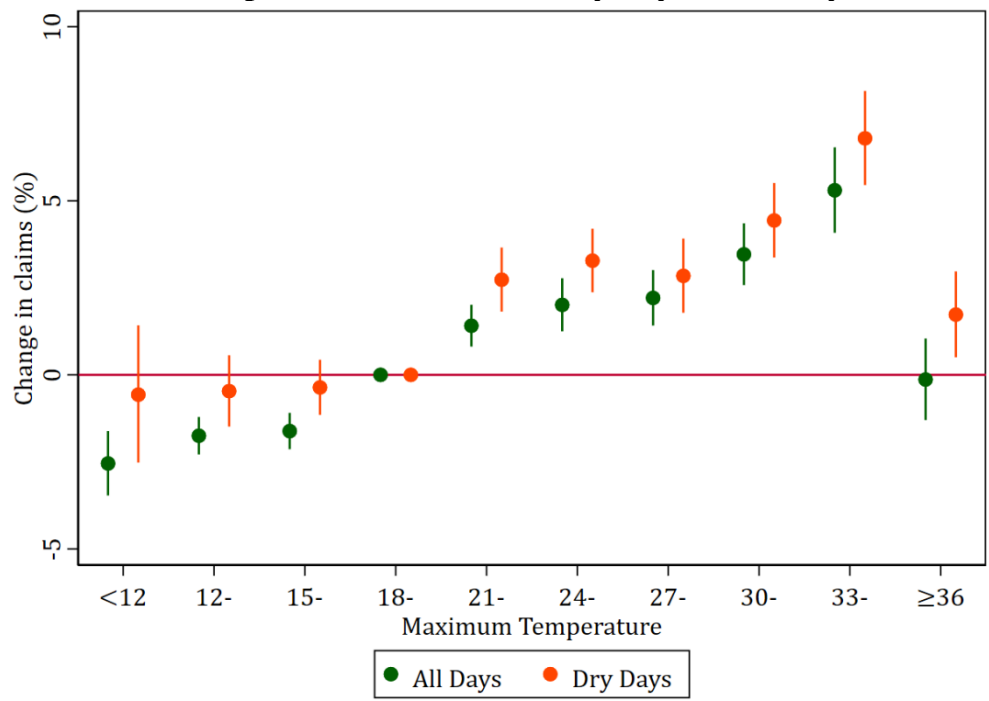

**Notes:** Estimated effects of daily maximum temperature (°C) for all occupations 1985–2020 using Poisson regression with robust standard errors clustered by postcode. Outcome variable is the number of claims in a postcode as outcome variable. semi-elasticity represents a percentage increase or decrease on the number of claims. 95% confidence intervals displayed.

**Figure E4. Maximum Temperature Estimates Including Weekends**

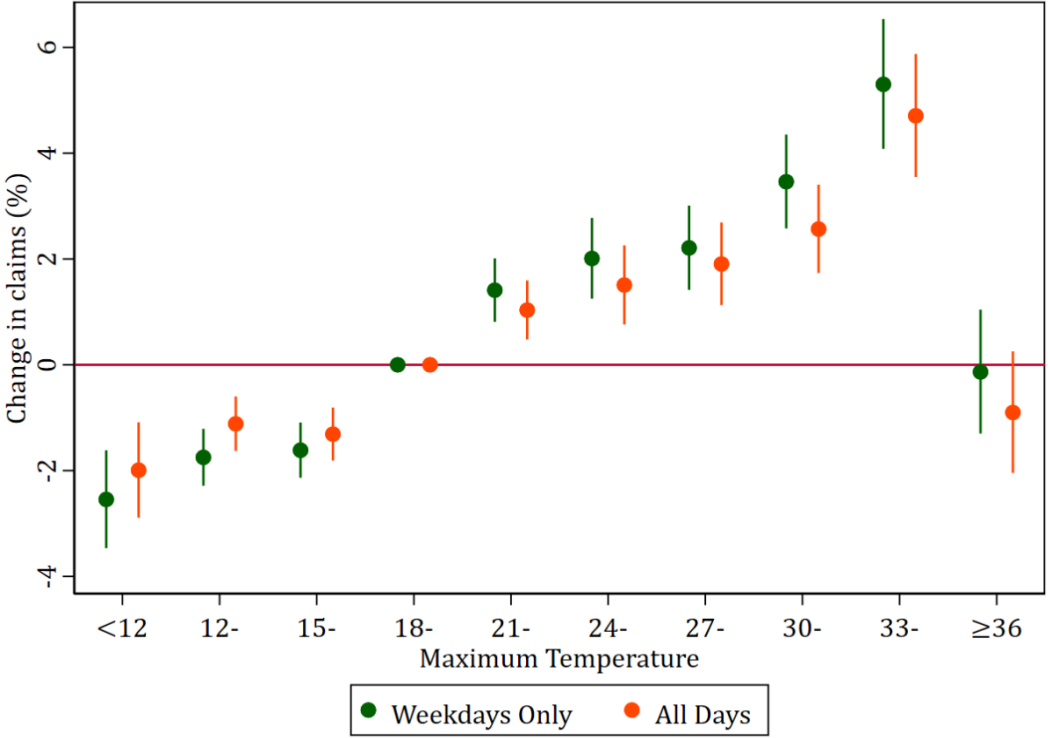

**Notes:** Estimated effects of daily maximum temperature (°C) for all occupations 1985–2020 using Poisson regression with robust standard errors clustered by postcode. Outcome variable is the number of claims in a postcode as outcome variable. 95% confidence intervals displayed.

Appendix F. Sample Size and Outcome Mean for Occupation Subgroup Regressions

**Table F1. Sample Size and Outcome Mean for Occupation Subgroup Regressions**

|  | **All Years** | **1985–1999** | **2000–2020** |
|---|---|---|---|
| All Occupations | μ=0.54<br>N=3,533,474<br>*p*=0.000 | μ=0.67<br>N=1,485,297<br>*p*=0.000 | μ=0.44<br>N=2,048,177<br>*p*=0.000 |
| High Risk Occupations | μ=0.33<br>N=2,882,065<br>*p*=0.000 | μ=0.46<br>N=1,278,261<br>*p*=0.000 | μ=0.24<br>N=1,603,804<br>*p*=0.000 |
| - *Tradespersons and related workers* | μ=0.21<br>N=2,063,608<br>*p*=0.000 | μ=0.27<br>N=870,614<br>*p*=0.000 | μ=0.16<br>N=1,192,994<br>*p*=0.000 |
| - *Labourers and related workers* | μ=0.23<br>N=2,350,818<br>*p*=0.000 | μ=0.31<br>N=1,141,102<br>*p*=0.000 | μ=0.15<br>N=1,209,716<br>*p*=0.000 |
| Other Occupations | μ=0.33<br>N=2,887,176<br>*p*=0.000 | μ=0.36<br>N=1,174,879<br>*p*=0.030 | μ=0.31<br>N=1,712,297<br>*p*=0.044 |
| - *Production and transport workers* | μ=0.18<br>N=1,669,468<br>*p*=0.005 | μ=0.19<br>N=723,530<br>*p*=0.122 | μ=0.17<br>N=945,938<br>*p*=0.119 |
| - *Intermediate clerical, sales and service workers* | μ=0.10<br>N=1,371,714<br>*p*=0.212 | μ=0.10<br>N=480,158<br>*p*=0.668 | μ=0.10<br>N=891,556<br>*p*=0.523 |
| - *Associate professionals* | μ=0.11<br>N=1,226,399<br>*p*=0.319 | μ=0.11<br>N=422,926<br>*p*=0.820 | μ=0.10<br>N=803,473<br>*p*=0.420 |
| - *Professionals* | μ=0.12<br>N=1,721,281<br>*p*=0.514 | μ=0.13<br>N=724,815<br>*p*=0.754 | μ=0.12<br>N=996,466<br>*p*=0.887 |
| - *Elementary clerical, sales and service workers* | μ=0.11<br>N=828,573<br>*p*=0.914 | μ=0.14<br>N=402,562<br>*p*=0.795 | μ=0.08<br>N=426,011<br>*p*=0.887 |
| - *Managers and administrators* | μ=0.07<br>N=830,111<br>*p*=0.914 | μ=0.08<br>N=388,405<br>*p*=0.540 | μ=0.07<br>N=441,706<br>*p*=0.887 |
| - *Advanced clerical and service workers* | μ=0.07<br>N=318,304<br>*p*=0.914 | μ=0.07<br>N=143,170<br>*p*=0.820 | μ=0.07<br>N=175,134<br>*p*=0.887 |

**Notes:** Outcome mean (μ), sample size (N) and Westfall-Young corrected p-values (p) for regressions in Table 3. The outcome mean is the average daily number of claims in a postcode. The sample size is the number of postcode-days in the regression. The Westfall and Young p-values adjust for multiple hypotheses testing within each panel (1000 replications). All regressions use Poisson Regression Month-Year-Postcode fixed-effects as per column (1) of Table 2; the inclusion of Month-Year-Postcode fixed-effects removes observations in postcode-months with no observations. Claims categorised using ASCO occupation codes. "Other" combines all claims for clerical, sales, service, production, transport, professional, associate professional, managers and administrators and workers into a single group.

Appendix G. Comparison of Main Results to U.S. Studies

We compare our results with two studies from the U.S. using occupational health claims from Texas (Dillender, 2021) and California (Park et al., 2021). These two studies are most suitable for comparison with our results due to their empirical approach with fixed -effects and large occupational health claims databases. For comparison, we convert estimates based on Fahrenheit to Celsius. We present results as percentage change in claims (or claims per 100,000 workers) for a given maximum temperature relative to day reaching 16–18°C.

Dillender analysis a sample 1,916,590 workers' compensation claims from Texas submitted between 2006 and 2014 using linear regression with fixed-effects. The dependant variable is daily claims per 100,000 workers. The "day-of-claim" results are used as the most suitable for comparison. Park et al. use inverse hyperbolic sine transformed regression with fixed-effects on a database of 11.1 million claims from 2001 to 2018 in California. They supress values in their results table, thus the entire distribution is not populated in Figure F1. We run our primary specification on a sample containing all occupations from 2000–2020 for a closer alignment. We use a baseline category of 15–18°C to more closely align to the estimates from the two studies.

**Figure G1. Comparison of Effect of Maximum Temperature on Occupational Health Claims**

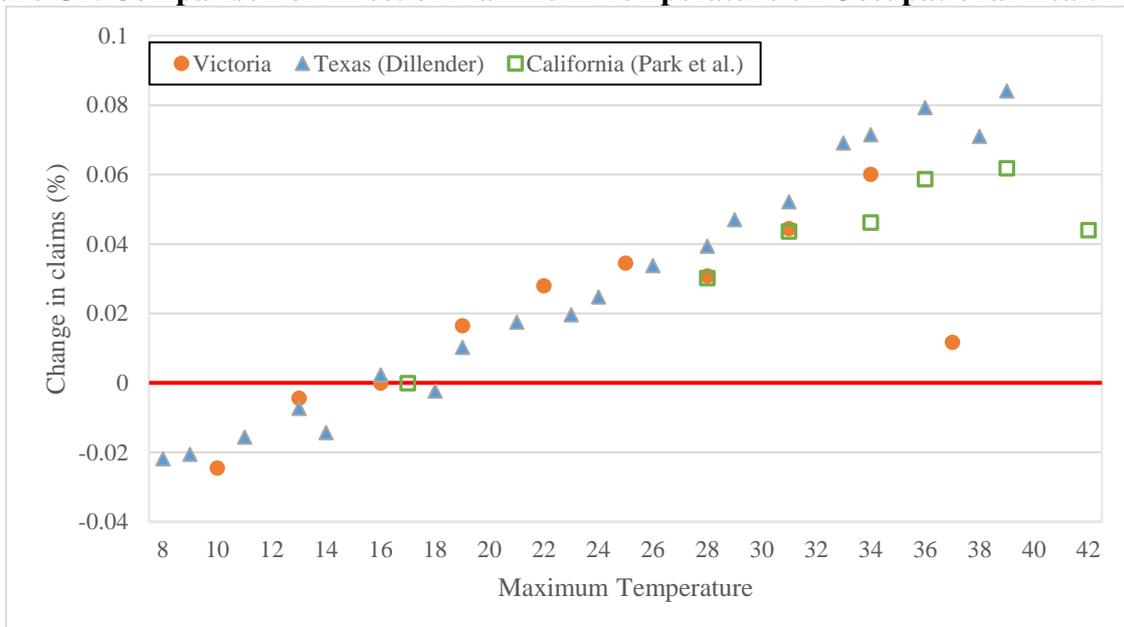

**Notes:** Estimated effects of maximum temperature on occupational health claims. Effects are displayed as a percentage change in the number of claims relative to a reference category of 16°C. The present study uses Poisson regression on claims from 2000–2020 in Victoria. Dillender, uses linear regression on claims from Texas between 2006 and 2014. Park et al. uses inverse hyperbolic sine transformed regression on claims from California from 2001–2018. All studies include regional and time related fixed-effects.

Appendix H. Effect of Heat by Industry

**Table H1. Maximum Temperature Estimate by Industry, Semi-elasticity (%)**

| Industry | Labourers and Tradespersons | Other Occupations |
|---|---|---|
| Agriculture, Forestry and Fishing | μ=0.06<br>N=221,120<br>*p*=0.043 | μ=0.05<br>N=84,501<br>*p*=0.966 |
| Accommodation and Food Services | μ=0.06<br>N=167,524<br>*p*=0.587 | μ=0.07<br>N=226,732<br>*p*=0.966 |
| Health Care and Social Assistance | μ=0.06<br>N=240,423<br>*p*=0.142 | μ=0.12<br>N=859,855<br>*p*=0.953 |
| Transport, Postal and Warehousing | μ=0.08<br>N=167,378<br>*p*=0.132 | μ=0.11<br>N=459,002<br>*p*=0.003 |
| Construction | μ=0.10<br>N=757,774<br>*p*=0.000 | μ=0.06<br>N=296,919<br>*p*=0.498 |
| Manufacturing | μ=0.270<br>N=727,092<br>*p*=0.000 | μ=0.13<br>N=537,43<br>*p*=0.911 |
| Administrative and Support Services | μ=0.08<br>N=231,631<br>*p*=0.270 | μ=0.07<br>N=135,988<br>*p*=0.953 |
| Wholesale Trade | μ=0.08<br>N=252,361<br>*p*=0.638 | μ=0.11<br>N=378,430<br>*p*=0.498 |
| Retail Trade | μ=0.06<br>N=235,415<br>*p*=0.665 | μ=0.08<br>N=434,467<br>*p*=0.953 |
| **All industries** | μ=0.24<br>N=1,603,804<br>*p*=0.00 | μ=0.31<br>N=1,712,297<br>*p*=0.000 |

**Notes:** Outcome mean (μ), sample size (N) and Westfall-Young corrected p-values (*p*) for regressions in Table 5. The outcome mean is the average daily number of claims in a postcode. The sample size is the number of postcode-days in the regression. The Westfall and Young p-values adjust for multiple hypotheses testing within each panel (1000 replications). All regressions use Poisson Regression Month-Year-Postcode fixed-effects as per column (1) of Table 2; the inclusion of Month-Year-Postcode fixed-effects removes observations in postcode-months with no observations.

**Table H2. Same-day Maximum Temperature Estimates by Industry, Semi-elasticity (%)**

| Industry | Labourers and Tradespersons | Other Occupations |
|---|---|---|
| Agriculture, Forestry and Fishing | 0.72 | 0.36 |
|  | (0.238) | (0.401) |
| Construction | 0.49 | 0.20 |
|  | (0.089) | (0.185) |
| Health Care and Social Assistance | 0.39 | 0.07 |
|  | (0.194) | (0.088) |
| Accommodation and Food Services | 0.36 | 0.01 |
|  | (0.278) | (0.179) |
| Manufacturing | 0.30 | 0.09 |
|  | (0.075) | (0.094) |
| Transport, Postal and Warehousing | 0.27 | 0.26 |
|  | (0.197) | (0.105) |
| Wholesale Trade | 0.16 | 0.06 |
|  | (0.179) | (0.123) |
| Administrative and Support Services | 0.12 | -0.14 |
|  | (0.180) | (0.249) |
| Retail Trade | 0.04 | 0.00 |
|  | (0.200) | (0.132) |
| **All industries** | 0.34 | 0.10 |
|  | (0.038) | (0.035) |

**Notes:** Estimates for effect of daily maximum temperature on number of claims using Poisson regression. Estimates displayed as a semi-elasticity, as a percentage increase or decrease in the number of claims. All regressions include Month-Year-Postcode fixed-effects as per column (1) of Table 2. Robust standard errors clustered at postcode level in parenthesis. 2000–2020. Industries with the most claims for labourers and tradespersons are displayed.

Appendix I: Sample Size and Outcome Mean for Worker Characteristic Regressions

**Table I1. Sample Size and Outcome Mean for Worker Characteristic Regressions**

|  | All Workers | Labourers and Tradespersons | Labourers and Tradespersons in Agriculture or Construction |
|---|---|---|---|
| Male | μ=0.33<br>N=1,819,392<br>*p*=0.000 | μ=0.21<br>N=1,472,897<br>*p*=0.000 | μ=0.10<br>N=1,019,784<br>*p*=0.000 |
| Female | μ=0.20<br>N= 1487427<br>*p*=0.528 | μ=0.09<br>N=730,564<br>*p*=0.127 | μ=0.05<br>N=123,746<br>*p*=0.052 |
| Age 18–35, % | μ=0.22<br>N=1,458,974<br>*p*=0.000 | μ=0.14<br>N=1,099,456<br>*p*=0.000 | μ=0.07<br>N=628,942<br>*p*=0.000 |
| Age 36–50, % | μ=0.22<br>N=1,521,111<br>*p*=0.034 | μ=0.13<br>N=1,001,266<br>*p*=0.001 | μ=0.07<br>N=514,076<br>*p*=0.027 |
| Age 51–64, % | μ=0.16<br>N=1,374,400<br>*p*=0.000 | μ=0.10<br>N=835,185<br>*p*=0.000 | μ=0.06<br>N=378,853<br>*p*=0.002 |
| Low earnings | μ=0.14<br>N=1,260,141<br>*p*=0.072 | μ=0.09<br>N=789,300<br>*p*=0.114 | μ=0.06<br>N=399,091<br>*p*=0.439 |
| Middle earnings | μ=0.15<br>N=1,192,181<br>*p*=0.024 | μ=0.09<br>N=768,299<br>*p*=0.090 | μ=0.06<br>N=374,804<br>*p*=0.040 |
| High earnings | μ=0.15<br>N=1,170,277<br>*p*=0.131 | μ=0.10<br>N=712,881<br>*p*=0.013 | μ=0.06<br>N=336,765<br>*p*=0.001 |
| Small | μ=0.13<br>N=1,425,827<br>*p*=0.000 | μ=0.10<br>N=1,067,920<br>*p*=0.000 | μ=0.07<br>N=739,759<br>*p*=0.018 |
| Medium | μ=0.26<br>N=1,428,141<br>*p*=0.000 | μ=0.17<br>N=1,022,702<br>*p*=0.000 | μ=0.09<br>N=465,989<br>*p*=0.000 |
| Large | μ=0.24<br>N=1,024,327<br>*p*=0.049 | μ=0.14<br>N=608,689<br>*p*=0.021 | μ=0.09<br>N=177,535<br>*p*=0.439 |
| All Claims | μ=0.44<br>N=2,048,177<br>*p*=0.000 | μ=0.24<br>N=1,603,804<br>*p*=0.000 | μ=0.09<br>N=932,163<br>*p*=0.000 |

**Notes:** Outcome mean (μ), sample size (N) and Westfall-Young corrected p-values (p) for regressions in Table 6. The outcome mean is the average daily number of claims in a postcode. The sample size is the number of postcode-days in the regression. The Westfall and Young p-values adjust for multiple hypotheses testing within each panel (1000 replications). All regressions use Poisson Regression Month-Year-Postcode fixed-effects as per column (1) of Table 2; the inclusion of Month-Year-Postcode fixed-effects removes observations in postcode-months with no observations. Low, medium and high wage based on terciles for the year and population of the respective column where provided. Firm size is based on total annual remuneration in 2010/11 deflated to 2005/06 dollars (<1Million = Small, 1–20 Million = Medium, >20 Million = Large).

**Table I2. Same-day Maximum Temperature by Worker Characteristic, Semi-elasticity (%)**

|  | All Workers | Labourers and Tradespersons | Labourers and Tradespersons in Agriculture or Construction |
|---|---|---|---|
| <u>Sex</u> | | | |
| Male | 0.27 (0.032) | 0.37 (0.046) | 0.44 (0.075) |
| Female | 0.06 (0.049) | 0.18 (0.100) | 0.73 (0.320) |
| | | | |
| <u>Age</u> | | | |
| Age 18–35, % | 0.25 (0.040) | 0.35 (0.058) | 0.53 (0.110) |
| Age 36–50, % | 0.16 (0.045) | 0.35 (0.075) | 0.31 (0.128) |
| Age 51–64, % | 0.20 (0.053) | 0.33 (0.092) | 0.60 (0.156) |
| | | | |
| <u>Earnings</u> | | | |
| Low earnings | 0.12 (0.068) | 0.22 (0.102) | 0.31 (0.171) |
| Middle earnings | 0.19 (0.056) | 0.15 (0.094) | 0.36 (0.169) |
| High earnings | 0.12 (0.069) | 0.28 (0.099) | 0.69 (0.182) |
| | | | |
| <u>Firm Size</u> | | | |
| Small | 0.22 (0.059) | 0.32 (0.082) | 0.43 (0.113) |
| Medium | 0.25 (0.043) | 0.35 (0.060) | 0.53 (0.113) |
| Large | 0.13 (0.049) | 0.23 (0.082) | 0.19 (0.193) |
| **All Claims** | 0.20 (0.026) | 0.34 (0.038) | 0.52 (0.086) |

**Notes:** Estimates for effect of daily maximum temperature on number of claims using Poisson regression for claims with afflictions occurring between 2000 and 2020. Estimates displayed as a semi-elasticity, as a percentage increase or decrease in the number of claims. All regressions include Month-Year-Postcode fixed-effects as per column (1) of Table 2. Low, medium and high wage based on terciles for the year and population of the respective column where provided. Robust standard errors clustered at postcode level in parenthesis. Firm size is based on total annual remuneration in 2010/11 deflated to 2005/06 dollars (<1Million = Small, 1–20 Million = Medium, >20 Million = Large).

Appendix J. Additional Details for Firm Analysis

**Table J1. Same-day Maximum Temperature on Occupational Health Claims by Firm Type**

|                       | "Low Claims" Firm | "High Claims" Firm |
|-----------------------|-------------------|--------------------|
| Maximum Temperature   | 0.70              | 0.20               |
|                       | (0.239)           | (0.149)            |
|                       |                   |                    |
| Number of firms       | 454               | 455                |
| Number of claims      | 7,729             | 33,084             |
| Outcome mean          | 0.07              | 0.11               |
| N (postcode-days)     | 111,514           | 296,694            |

**Notes:** Based on claims for labourers and tradespersons occurring between 2005–2019 for a subset of firms in Construction, Manufacturing and Agriculture, Forestry & Fishing. Firms are classified as "high claims" and "low claims" firms compared with other firms in the same industry and size using the individual firm fixed-effect (firms above the median considered "high claims"). Regressions include Month-Year-Postcode fixed-effects as per column (1) of Table 2. Robust standard errors clustered at postcode level in parenthesis.

Appendix K. Calculations for Cost of Heat on Workers' Health

**Table K1. Distribution of Maximum Temperature in Victoria**

| Maximum Temperature | Average Days P/A | | Δ Days P/A |
|---|---|---|---|
| | Sept'85–Aug'90 | Aug'15–Jul'20 | |
| Less than 12°C | 17.8 | 12.6 | -5.2 |
| 12–15°C | 78.8 | 75 | -3.8 |
| 15–18°C | 72.8 | 68.4 | -4.4 |
| 18–21°C | 69.4 | 56.6 | -12.8 |
| 21–24°C | 49.4 | 50.6 | 1.2 |
| 24–27°C | 32.2 | 38 | 5.8 |
| 27–30°C | 19.6 | 28.4 | 8.8 |
| 30–33°C | 14 | 17.4 | 3.4 |
| 33–36°C | 6.4 | 9.8 | 3.4 |
| Greater than or equal 36°C | 4.8 | 8.6 | 3.8 |

**Notes:** Annual distribution of daily maximum temperature of Victorian postcodes weighted by the area's number of claims in the CRD.

**Table K2. Annual Cost of Change in Temperature Distribution on Occupational Health**

| | Δ Days P/A | Claims | | All Injuries/Diseases | |
|---|---|---|---|---|---|
| | | Daily Δ in claims | Cost P/A (AU$ mil.) | Daily Δ in cases | Cost P/A (AU$ mil.) |
| *Panel A. All Workers* | | | | | |
| Less than 12°C | -5.2 | -2.9 | 3.9 | -8.0 | 5.3 |
| 12–15°C | -3.8 | -2.0 | 2.0 | -5.5 | 2.7 |
| 15–18°C | -4.4 | -1.8 | 2.1 | -5.1 | 2.8 |
| 18–21°C | -12.8 | 0.0 | .0 | 0.0 | .0 |
| 21–24°C | 1.2 | 1.6 | .5 | 4.5 | .7 |
| 24–27°C | 5.8 | 2.3 | 3.4 | 6.4 | 4.7 |
| 27–30°C | 8.8 | 2.5 | 5.7 | 7.0 | 7.8 |
| 30–33°C | 3.4 | 3.9 | 3.5 | 10.9 | 4.7 |
| 33–36°C | 3.4 | 6.0 | 5.3 | 16.8 | 7.2 |
| Greater than or equal 36°C | 3.8 | -0.2 | -.2 | -0.4 | -0.2 |
| TOTAL | - | 9.5 | 26.3 | 26.4 | 35.6 |
| *Panel B. Labourers and Tradespersons* | | | | | |
| Less than 12°C | -5.2 | -1.4 | 1.9 | -3.9 | 2.6 |
| 12–15°C | -3.8 | -1.1 | 1.1 | -3.2 | 1.5 |
| 15–18°C | -4.4 | -0.9 | 1.1 | -2.6 | 1.5 |
| 18–21°C | -12.8 | 0.0 | .0 | 0.0 | .0 |
| 21–24°C | 1.2 | 0.9 | .3 | 2.6 | .4 |
| 24–27°C | 5.8 | 1.3 | 2.0 | 3.7 | 2.7 |
| 27–30°C | 8.8 | 1.5 | 3.4 | 4.1 | 4.6 |
| 30–33°C | 3.4 | 2.3 | 2.1 | 6.5 | 2.8 |
| 33–36°C | 3.4 | 3.4 | 3.0 | 9.5 | 4.1 |
| Greater than or equal 36°C | 3.8 | 0.8 | .8 | 2.2 | 1.1 |
| TOTAL | - | 6.8 | 15.6 | 18.9 | 21.2 |

**Notes:** Daily impact of number of claims/cases is calculated using the semi-elasticities of same-day maximum temperature for 1985-2020. Cost in Australia dollars using unit costs of AU$258700 for compensated claims and AU$126,500 for all injuries and diseases. The daily number of injuries and diseases is based on the FY12/13 (115,415 injuries and diseases and 41,586 claims).